\documentclass[a4paper,12pt]{article}
\include{epsf}
\usepackage{latexsym}
\usepackage{amssymb}
\usepackage{amsmath}
\usepackage[dvips]{graphicx}
\newlength{\cqfd}
\def\R{\mathbb{R}}

\newtheorem{theorem}{Theorem}

\newcommand{\promille}{%
  \relax\ifmmode\promillezeichen
        \else\leavevmode\(\mathsurround=0pt\promillezeichen\)\fi}
\newcommand{\promillezeichen}{%
  \kern-.05em%
  \raise.5ex\hbox{\the\scriptfont0 0}%
  \kern-.15em/\kern-.15em%
  \lower.25ex\hbox{\the\scriptfont0 00}}

\title{\bf Roll-waves in bi-layer flows}

\begin{document}
 \maketitle 
\begin{center}
{\large Marc Boutounet  \footnote{ONERA, 2 Avenue E. Belin
31055 Toulouse Cedex, France; boutounet@insa-toulouse.fr} 
Pascal Noble \footnote{ Universit\'e de Lyon, Universit\'e Lyon 1
Institut Camille Jordan, UMR CNRS 5208
 43, blvd du 11 novembre 1918,
F - 69622 Villeurbanne Cedex, France; noble@math.univ-lyon1.fr:
Research of P.N. was partially supported by French ANR project
no. ANR-09-JCJC-0103-01} 
Jean-Paul Vila \footnote{Institut de Math\'ematiques de Toulouse, UMR CNRS 5219,
INSA de Toulouse, 135 avenue de Rangueil, 31077 Toulouse Cedex 4 - France; vila@insa-toulouse.fr}}
\end{center}

\vspace {0.5cm}
\begin{center}
{\bf Abstract.}
\end{center}
In this paper, we derive consistent shallow water equations for
bi-layer flows of Newtonian fluids flowing down a ramp. We carry out a
complete spectral analysis of steady flows in the low frequency regime
and show the occurence of hydrodynamic instabilities, so called 
roll-waves, when steady flows are unstable.

\section{Introduction}

This paper is devoted to the analysis of the gravity driven motion of a
superposition of two immiscible Newtonian fluids flowing down an
inclined plane.  Such systems can describe a lot of situations in
geophysics and engineering: mud flows, submarine avalanches, transport
of mass, heat and momentum in chemical technology, coating layers in
photography. For this latter application, the formation of waves is
highly indesirable. It is then an important problem to study the
stability of such multiple-layer flows. Linear stability analysis was
addressed in many papers: see e.g. \cite{Kao1},\cite{Kao2},\cite{Koba}.  
But these studies did not provide a model to describe the nonlinear waves.
Indeed, modeling such systems is a hard problem both from the
mathematical and numerical viewpoint: in particular, one has to deal
with two free surfaces: one at the fluid interface and the other one 
at the interface between fluid and gas.\\

\noindent
Here, we consider the particular situation where two
{\it thin} fluids are flowing down a ramp. This means that the
characteristic depths of the fluids are much smaller than the 
characteristic length of the flow in the downstream direction. We 
take advantage of the thinness of the layers to write a reduced system
of equations which will contain all the physical ingredients that are 
relevant to describe the dynamics of such flows. A similar strategy
was developed by Kliakhandler in \cite{kliak}: a system of Kuramoto
Sivashinsky equations is derived from the full Navier Stokes equations 
in the presence of surface tension. Using this approach, the author
analysed the spectral stability of two-layered thin film
flows and considered in particular the interaction between convection
and each relevant physical term: buoyancy, inertia and capillarity. In
particular, it is proved that in some parameter regime, the convection
can stabilize an unstable density stratification. Though this spectral
analysis highlights the role of each term (buoyancy, inertia,
capillarity), it is not complete since one has to consider the
interaction between  {\it all} the relevant terms. In particular, the
competition between inertia and buoyancy is the source of
hydrodynamical instabilities in shallow waters. Moreover, 
the derivation of Kuramoto Sivashnisky equations is usually limited 
to small amplitude motions. The purpose of this paper is to obtain a 
system of shallow water equations which is consistent with
Navier-Stokes equations in the regime
of shallow waters. In order to derive such a system, we  
follow the methodology introduced by Vila \cite{Vila} and justified 
rigorously by Bresch and Noble \cite{Bresch_Noble} for a single fluid
layer. As a byproduct, the system of shallow water equations is
relevant to study the linear stability of steady flows in the low
frequency regime. We will complete the spectral analysis of \cite{kliak} 
in some particular cases (stable/unstable mass stratification, viscous
stratification). We prove that the system of Kuramato
Sivashinsky equations in \cite{kliak} is also contained in our
model in some specific regimes. Finally, we use shallow water
equations to describe nonlinear waves when steady flows are
spectrally unstable. In particular, we show the occurence of well
known hydrodynamic instabilities, so called roll-waves, which can
appear either at fluid interface and free surface or only at the fluid
interface.\\

The paper is organized as follows. In section \ref{sec1}, we describe
bi-layer flows in the shallow water scaling and compute an expansion
of the velocity and pressure field in this regime. From
this asymptotic analysis, we find the system of Kuramoto Sivashinsky 
equations of \cite{kliak} and study the spectral stability of steady 
flows in the low frequency regime. Then we derive shallow water equations. In
section \ref{sec2}, we prove the existence of small amplitude
roll-waves when the steady flow is unstable. Finally, we compute
numerically large amplitude roll-waves through direct numerical
simulations of the shallow water equations.

\section{\label{sec1} Shallow Water Eqs. for bi-layer flows}

In this section, we show how to expand solutions to Navier-Stokes
equations in the regime of shallow water. With these expansions, we
obtain a hierarchy of models for bi-layer shallow flows. First, we
write lubrication models: using zeroth (resp. first order) expansion of
the velocity field, we obtain a system of inviscid (resp. viscous)
conservation laws on the fluid heights. We find the system of coupled
Kuramoto-Sivashinsky equations in \cite{kliak} if we take into account 
of the capillary forces. We use this system of equations to study the
spectral stability of steady states in the low frequency regime. Next, 
using first order expansions of the fluids velocities, we derive
inviscid shallow water equations. For this latter step, one has to
carry out a closure procedure: we have chosen to write the tangential
stress at the bottom and at the interface proportionnal respectively
to the average velocity at the bottom and the difference between
average fluid velocities (up to correction terms).

\subsection{Description of bi-layer flows in the shallow water regime}

In this part, we write Navier-Stokes equations for bi-layer flows in a
nondimensional form in the regime of shallow waters. We then perform
an asymptotic expansions of solutions with respect to the so-called
{\it aspect ratio} (defined hereafter) in the neighbourhood of a
Nusselt steady solution.

\subsubsection{Scaling Navier-Stokes equations}
Let us consider the superposition of two incompressible and immiscible
fluids with density, viscosity and capillarity $(\rho_i,\nu_i,\sigma_i), i=1,2$ flowing
down an inclined plane with a slope $\theta$ (see figure \ref{fig1}).
\begin{figure}[h!]
\begin{center}
\input{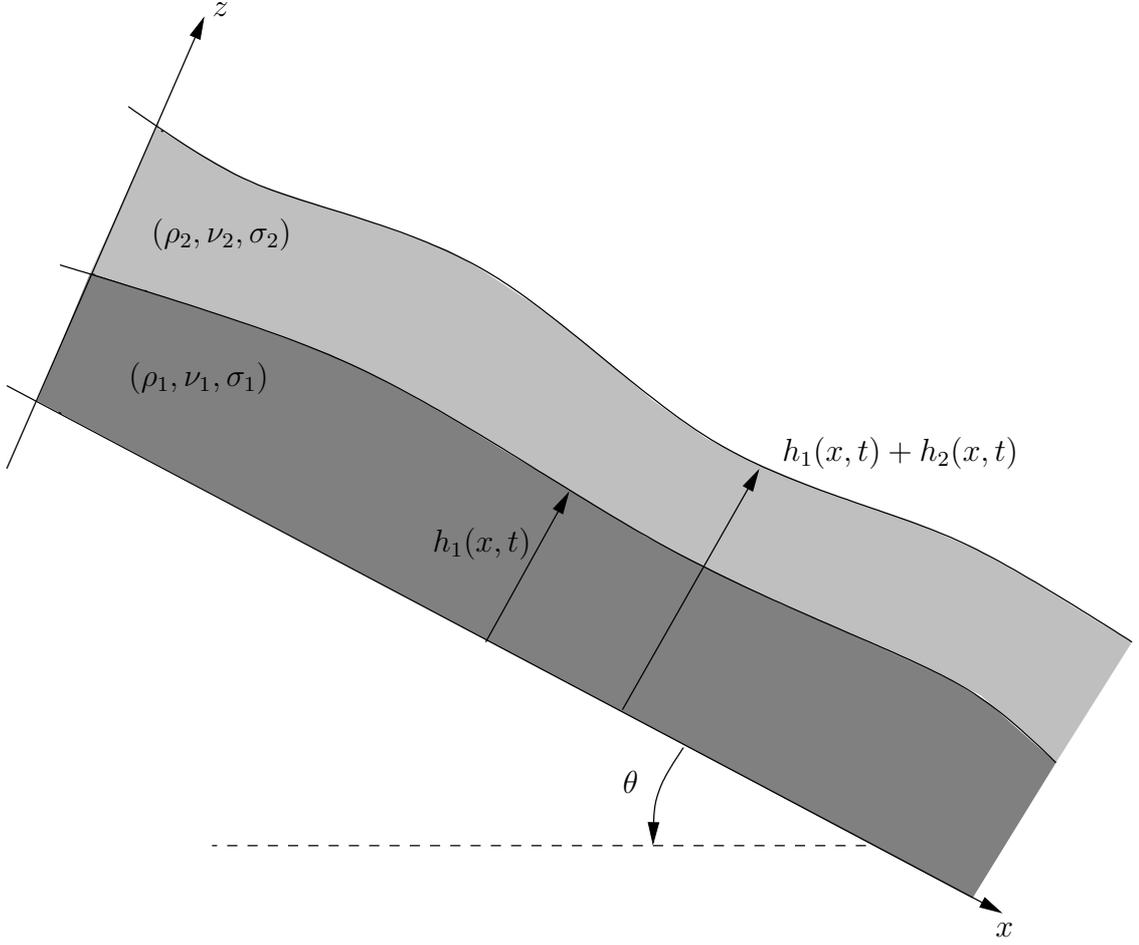}
\end{center}
\caption{\label{fig1} Two fluids flowing down an inclined plane.}
\end{figure}
\noindent
We introduce the aspect ratio $\varepsilon$, the Reynolds number
$R_e$, the Froude number $F$ and Weber numbers $W_i$ as 
\begin{equation*}
\displaystyle
\varepsilon=\frac{H}{L},\quad R_e=\frac{\rho_1 HU}{\nu_1},\quad
F^2=\frac{U^2}{gH},\quad W_i=\frac{\sigma_i}{\rho_i HU^2},\:i=1,2,
\end{equation*}
where $H$ denotes the characterictic depth of the fluid and $L$ the
characteristic length in the streamwise direction. The characteristic
fluid velocity $U$ can be chosen as the average velocity in the fluid
layer for a Nusselt flow. We further introduce the additionnal numbers
\begin{equation*}
\displaystyle
\rho=\frac{\rho_2}{\rho_1},\quad \nu=\frac{\nu_2}{\nu_1}.
\end{equation*}  
\noindent
The motion of fluids (1) and (2) is described by Navier-Stokes equations
{\setlength\arraycolsep{1pt}
\begin{eqnarray}
\displaystyle
\varrho_i\big(\partial_t u_i+\partial_x u_i^2+\partial_z u_iw_i\big)+\frac{\partial_x p_i}{F^2}&=&\frac{s\varrho_i}{\varepsilon F^2}+\frac{\mu_i}{\varepsilon R_e}\big(\partial_{zz}u_i+\varepsilon^2\partial_{xx}u_i\big),\\
\displaystyle
\varrho_i\big(\partial_t w_i+\partial_x u_iw_i+\partial_z w_i^2\big)+\frac{\partial_zp_i}{\varepsilon^2 F^2}&=&-\frac{\varrho_i c}{\varepsilon^2 F^2}+\frac{\mu_i}{\varepsilon R_e}\big(\partial_{zz}w_i+\varepsilon^2\partial_{xx}w_i\big),\\
\displaystyle
\partial_x u_i+\partial_z w_i&=&0,\quad i=1,2.
\end{eqnarray}}
Here $\varrho_1=1$, $\varrho_2=\rho$, $\mu_1=1$, $\mu_2=\nu$. These
equations are set in the fluid
domains $$\Omega_{1,t}=\left\{(x,z)\in\mathbb{R}^2/0\leq z\leq
  h_1(x,t)\right\}$$
and $$\Omega_{2,t}=\left\{(x,z)\in\mathbb{R}^2/h_1(x,t)\leq z\leq
  h_1+h_2(x,t)=h(x,t)\right\}.$$ The kinematic conditions at the
bottom, fluids interface and free surface are respectively
\begin{equation}\label{kin_eq}
\begin{array}{lll}
\displaystyle
u_1(0,x)=w_1(0,x)=0,\quad u_1(h_1)=u_2(h_1),\quad w_1(h_1)=w_2(h_1),\\
\\
\displaystyle
\partial_t h_1+u_1(h_1)\partial_x h_1=w_1(h_1),\quad \partial_t h+u_2(h)\partial_x(h)=w_2(h).
\end{array}
\end{equation}
\noindent
We assume the continuity of the fluid stress at the fluids interface
and at the free surface. First, the continuity of normal stresses yields
{\setlength\arraycolsep{1pt}
\begin{eqnarray*}
\displaystyle
p_2(h)&=&-\frac{\kappa_2\,F^2\partial_{xx}h}{\big(1+\varepsilon^2(\partial_x
  h)^2\big)^{\frac{3}{2}}}-\frac{2\nu\varepsilon F^2}{R_e}\partial_xu_2(h)\frac{1+\varepsilon^2(\partial_x h)^2}{1-\varepsilon^2(\partial_x h)^2},\\
 \displaystyle
p_1(h_1)-p_2(h_1)&=&-\frac{\kappa_1\,F^2\partial_{xx}h_1}{\big(1+\varepsilon^2(\partial_x
  h_1)^2\big)^{\frac{3}{2}}}\nonumber\\
\displaystyle
&&-\frac{2\varepsilon F^2}{Re}\big(\partial_xu_1(h_1)-\nu\partial_x u_2(h_1)\big)\frac{1+\varepsilon^2(\partial_x h_1)^2}{1-\varepsilon^2(\partial_x h_1)^2},
\end{eqnarray*}}

\noindent
with $\kappa_i=\varepsilon^2\,W_i$. In order to take into account of
the surface tension effects, we assume $\kappa_i=\mathcal{O}(1)$. Next, the continuity of tangential stresses yields 
{\setlength\arraycolsep{1pt}
\begin{eqnarray*}
\displaystyle
\big(\partial_z u_2&+&\varepsilon^2\partial_x
w_2\big)(h)=4\varepsilon^2\frac{\partial_x u_2(h)}{1-\varepsilon^2\big(\partial_x h\big)^2}\partial_x h,\nonumber\\
\displaystyle
\nu\big(\partial_z u_2&+&\varepsilon^2\partial_x w_2\big)(h_1)-\big(\partial_z u_1+\varepsilon^2\partial_x w_1\big)(h_1)=4\varepsilon^2\frac{\big(\nu\partial_x u_2-\partial_x u_1\big)(h_1)}{1-\varepsilon^2(\partial_x h_1)^2}\partial_x h_1.
\end{eqnarray*}}

Let us now describe the stationary solutions of this system. The velocity field does not depend on $x$ and $t$. The fluid heights are constant $h_1(x,t)=\overline{h}$, $h_2(x,t)=1-\overline{h}$ whereas the pressure is hydrostatic
\begin{equation*}
\displaystyle
p_1(z)=c(\overline{h}-z)+\rho c(1-\overline{h}),\:\:\forall 0\leq z\leq \overline{h},\quad p_2(z)=\rho c(1-z),
\end{equation*}
and the fluid velocities have a parabolic profile
\begin{equation*}
\begin{array}{ll}
\displaystyle
u_1(z)=\lambda\big(\rho(1-\overline{h})z+\overline{h} z-\frac{z^2}{2}\big),\quad\forall 0\leq z\leq \overline{h},\\ 
\displaystyle
u_2(z)=\lambda \overline{h}(\rho (1-\overline{h})+\frac{\overline{h}}{2})+\frac{\lambda\rho}{\nu}\Big((1-\overline{h})(z-\overline{h})-\frac{(z-\overline{h})^2}{2}\Big),\quad \forall \overline{h}\leq z\leq 1,
\end{array}
\end{equation*}
where $\lambda$ is a constant defined as $\displaystyle\lambda=\frac{Re\sin\theta}{F^2}$.\\

\noindent
In what follows, we will analyse the bi-layer flows in the neighbourhood of such steady solutions: this yields a natural scale for the characteristic fluid velocity $U$ and thus the constant $\lambda$ has to satisfy an extra relation. If one choose the ratio between the total mass discharge rate and the total mass of the fluid then  
\begin{equation*}
\begin{array}{ll}
\displaystyle
\int_0^{\overline{h}}u_1+\rho\int_{\overline{h}}^1 u_2=\overline{h}+\rho(1-\overline{h}),\\
\displaystyle
\lambda=3\frac{\overline{h}+\rho(1-\overline{h})}{\displaystyle\overline{h}^3+3\rho\overline{h}^2(1-\overline{h})+3\rho^2\overline{h}(1-\overline{h})^2+\frac{\rho^2}{\nu}(1-\overline{h})^3}.
\end{array}
\end{equation*}
\noindent
Note that for a single layer of fluid ($\rho=\nu=1$), one recovers the condition
of Vila $\lambda=3$. Another possible choice for the characteristic
velocity would be the fluid velocity at the free surface: one then
recovers the classical value $\lambda=2$. In both cases, there is a
relation between the Reynolds and Froude numbers. Here, we
have chosen $U$ so that $\lambda=3$: 
there remains $R_e$, $\theta$, $\overline{h}$, $\kappa_i$, $\rho,\nu$ 
as free parameters to design an experiment and describe a bi-layer
flow of Newtonian fluids.\\
  
\noindent
We will derive shallow water equations from Navier Stokes equations
integrated across each fluid layer (see e.g. \cite{shka}, \cite{r-quil1}, \cite{perthame}). First, we integrate the divergence free conditions on each fluid layer: using the kinematic conditions (\ref{kin_eq}), we find the mass conservation laws:
\begin{equation*}
\displaystyle
\partial_t h_1+\partial_x\big(\int_0^{h_1}u_1(z)dz\big)=0,\quad \partial_t h_2+\partial_x\big(\int_{h_1}^{h}u_2(z)dz\big)=0.
\end{equation*}
Denote $q_1=h_1\overline{u}_1=\int_0^{h_1}u_1$ and $q_2=h_2\overline{u}_2=\int_{h_1}^{h}u_2$ the discharge rates in the streamwise direction: the mass conservation laws then read
\begin{equation}\label{cons_mass0}
\displaystyle
\partial_t h_1+\partial_x(h_1\overline{u}_1)=0,\quad \partial_t h_2+\partial_x(h_2\overline{u}_2)=0.
\end{equation}
\noindent
Now, we write  a system of equations which governs the evolution of
$q_i=h_i\overline{u}_i$.  This is done through the integration of momentum equations across each fluid layer:
{\setlength\arraycolsep{1pt}
\begin{eqnarray}\label{av_eq_mom}
\displaystyle
\partial_t\big(\int_0^{h_1}u_1\big)+\partial_x\Big(\int_0^{h_1}u_1^2&+&\frac{p_1}{F^2}\Big)+\frac{\kappa_1\partial_x
  h_1\partial_{xx}h_1}{\big(1+\varepsilon^2(\partial_x h_1)^2\big)^{\frac{3}{2}}}=-\frac{\partial_z u_1(0)}{\varepsilon R_e}\nonumber\\
\displaystyle
&&\frac{\lambda}{\varepsilon R_e}h_1+\frac{2\varepsilon}{R_e}\partial_x\Big(\int_0^{h_1}\partial_x u_1\Big)-\mathcal{T}\nonumber\\
\displaystyle
\partial_t\big(\rho\int_{h_1}^{h}u_2\big)+\partial_x\Big(\int_{h_1}^{h}\rho\,u_2^2&+&\frac{p_2}{F^2}\Big)+\frac{\kappa_2\partial_x
  h\partial_{xx}h}{\big(1+\varepsilon^2(\partial_x h)^2\big)^{\frac{3}{2}}}=\nonumber\\
\displaystyle
&&\frac{\lambda\rho h_2}{\varepsilon R_e}+\frac{2\varepsilon}{R_e}\partial_x\Big(\int_{h_1}^h\partial_x u_2\Big)+\mathcal{T},
\end{eqnarray}}
with $\mathcal{T}$ defined as 
\begin{equation*}
\displaystyle
\mathcal{T}=-\frac{p_2(h_1)\partial_x h_1}{F^2}-\frac{\nu}{\varepsilon R_e}\partial_z u_2(h_1)+\frac{\nu\varepsilon}{R_e}\big(2\partial_x u_2(h_1)\partial_x h_1-\partial_x w_2(h_1)\big).
\end{equation*}
\noindent
In order to write this evolution system in a closed form, one has to
find a relation between the different integrated quantities, the
tangential stresses at the wall, at the fluids interface,
$\mathcal{T}$ and the unknowns $h_i,q_i$. We follow the
method introduced by Vila \cite{Vila} in the case of a single
fluid layer. We expand the velocity field with respect to
$\varepsilon$ in order to find an expansion of the above quantities
and $q_i$ as functions of $h_i$ and their derivatives to any fixed
order. For a given order, this enables us to write the unknown
quantities in system (\ref{av_eq_mom}) as functions of $(h_i,q_i)$ and
derive a shallow water model in a closed form. 

\subsubsection{Asymptotic expansions of solutions to Navier-Stokes eqs}

In the shallow water regime $\varepsilon\approx 0$, the fluid
velocities and pressures $u_i,w_i,p_i$ almost satisfy a differential
system in $z$. The ``horizontal'' fluid velocities $u_i, i=1,2$ are solution to:
\begin{equation}
\displaystyle
\mu_i\partial_{zz}u_i+\varrho_i\lambda=\varepsilon R_e\varrho_i\big(\partial_t u_i+u_i\partial_x u_i+w_i\partial_z u_i \big)+\frac{\varepsilon R_e}{F^2}\partial_x p_i-\mu_i\varepsilon^2\partial_{xx}u_i.
\end{equation}
\noindent
We add the boundary conditions:
{\setlength\arraycolsep{1pt}
\begin{eqnarray*}
\displaystyle
\partial_z u_2(h)&=&4\varepsilon^2\frac{\partial_x u_2(h)}{1-\varepsilon^2\big(\partial_x h\big)^2}\partial_x h-\varepsilon^2\partial_x w_2(h),\nonumber\\
\displaystyle
\nu\partial_z u_2(h_1)-\partial_z u_1(h_1)&=&4\varepsilon^2\frac{\big(\nu\partial_x u_2-\partial_x u_1\big)(h_1)}{1-\varepsilon^2(\partial_x h_1)^2}\partial_x h_1\nonumber\\
\displaystyle
&&-\varepsilon^2\big(\nu\partial_x w_2(h_1)-\partial_x w_1(h_1)\big),
\end{eqnarray*}}
\noindent
and $\displaystyle u_1(0)=0,\quad u_1(h_1)=u_2(h_1)$.\\

\noindent
The fluid pressures are solutions to the differential system
\begin{equation}
\displaystyle
\partial_z p_i+\varrho_i c=\mu_i\frac{\varepsilon F^2}{R_e}\partial_{zz}w_i-\varepsilon^2F^2\varrho_i\Big(\partial_t w_i+u_i\partial_x w_i+w_i\partial_z w_i\Big)+\mu_i\frac{\varepsilon^3F^2}{Re}\partial_{xx}w_i,
\end{equation}
\noindent
whereas the boundary conditions for this system are given by
{\setlength\arraycolsep{1pt}
\begin{eqnarray*}
\displaystyle
p_2(h)&=&-\frac{\kappa_2\,F^2\partial_{xx}h}{\big(1+\varepsilon^2(\partial_x
  h)^2\big)^{\frac{3}{2}}}-\frac{2\nu\varepsilon F^2}{R_e}\partial_xu_2(h)\frac{1+\varepsilon^2(\partial_x h)^2}{1-\varepsilon^2(\partial_x h)^2},\\
 \displaystyle
p_1(h_1)-p_2(h_1)&=&-\frac{\kappa_1\,F^2\partial_{xx}h_1}{\big(1+\varepsilon^2(\partial_x
  h_1)^2\big)^{\frac{3}{2}}}\nonumber\\
\displaystyle
&&-\frac{2\varepsilon F^2}{Re}\big(\partial_xu_1(h_1)-\nu\partial_x u_2(h_1)\big)\frac{1+\varepsilon^2(\partial_x h_1)^2}{1-\varepsilon^2(\partial_x h_1)^2},
\end{eqnarray*}}
\noindent
Finally, the vertical velocities are solutions to
\begin{equation*}
\displaystyle
\partial_z w_i=-\partial_x u_i,\quad w_1(0)=0,\quad w_1(h_1)=w_2(h_1).
\end{equation*}
\noindent
There are three nondimensional numbers that are relevant to
parametrize this set of equations: let us define 
\begin{equation}
\displaystyle
\alpha=\frac{\varepsilon F^2}{R_e},\quad \beta=\varepsilon R_e,\quad \delta=\frac{\varepsilon R_e}{F^2}.
\end{equation}
\noindent
In what follows, we will assume that $\alpha,\beta,\delta\ll 1$ so as
to remain close to Nusselt type solutions. These assumptions are
clearly satisfied when $R_e, F=O(1)$ but a wider ranger of parameters
is valid. Next, we compute an Hilbert expansion of the fluid velocity and pressure:
\begin{equation*}
\displaystyle
u_i=\sum_{k=0}^{\infty}u_i^{(k)},\quad p_i=\sum_{k=0}^{\infty}p_i^{(k)}, 
\end{equation*}
so that
\begin{equation*}
\displaystyle
u_i-\sum_{k=0}^{j}u_i^{(k)}=\mathcal{O}\Big((\alpha+\beta+\delta)^{j+1}\Big),\quad p_i-\sum_{k=0}^{j}p_i^{(k)}=\mathcal{O}\Big((\alpha+\beta+\delta)^{j+1}\Big).
\end{equation*}
\noindent
Let us first compute $u_i^{(0)}, p_i^{(0)}$ Letting
$\alpha,\beta,\delta \to 0$ in the above equations leads a
differential system in the $z$ variable that is similar to the one
which determines stationary solutions. The fluid pressure is (up to this order) hydrostatic:
\begin{equation}
\begin{array}{ll}
\displaystyle
p_1^{(0)}(z)=c(\rho h_2+h_1-z)-\kappa_1F^2\partial_{xx}h_1-\kappa_2F^2\partial_{xx} h,\\
\displaystyle
p_2^{(0)}(z)=\rho c(h_1+h_2-z)-\kappa_2F^2\partial_{xx}h.
\end{array}
\end{equation}
\noindent
The fluid velocities in the streamwise direction have a parabolic profile
\begin{equation}
\begin{array}{ll}
\displaystyle
u_1^{(0)}(z)=\lambda\big(\rho h_2 z+h_1 z-\frac{z^2}{2}\big),\\
\displaystyle
u_2^{(0)}(z)=\lambda h_1\big(\rho h_2+\frac{h_1}{2}\big)+\frac{\lambda\rho}{\nu}\big(h_2(z-h_1)-\frac{(z-h_1)^2}{2}\big).
\end{array}
\end{equation}
The computation of higher order terms is then straightforward: assume
that we have computed $u_i^{(j)},p_i^{(j)},\;j\leq k$, then
$u_i^{(k+1)}$ is calculated by computing the solution to 
\begin{equation}
\displaystyle
\mu_i\partial_{zz}u_i^{(k+1)}=F_{i,k}(u_n^{(j)},p_n^{(j)}),\quad j\leq k,\quad n=1,2
\end{equation}
 with the boundary conditions
\begin{equation}
\begin{array}{ll}
\displaystyle
\partial_z u_2^{(k+1)}(h_1+h_2)=g_2^{(k)},\quad \nu\partial_z u_2^{(k+1)}(h_1)-\partial_z u_1^{(k+1)}(h_1)=g_1^{(k)},\\
\displaystyle
u_1^{(k+1)}(h_1)=u_2^{(k+1)}(h_1),\quad u_1^{(k+1)}(0)=0.
\end{array}
\end{equation}
\noindent
The solution $u_i^{(k+1)}$ to this system is  
{\setlength\arraycolsep{1pt}
\begin{eqnarray*}
\displaystyle
u_1^{(k+1)}&=&z\Big(\nu\,g_2^{(k)}-g_1^{(k)}-\int_{h_1}^{h}F_{2,k}(y)dy\Big)-\int_0^z\int_{\overline{z}}{h_1}F_{1,k}(y)dy\,d\overline{z},\\
\displaystyle
u_2^{(k+1)}&=&h_1\Big(\nu\,g_2^{(k)}-g_1^{(k)}-\int_{h_1}^{h}F_{2,k}(y)dy\Big)-\int_0^{h_1}\int_{\overline{z}}^{h_1}F_{1,k}(y)dyd\overline{z}\\
\displaystyle
&+&g_2^{(k)}(z-h_1)-\frac{1}{\nu}\int_{h_1}^z\int_{\overline{z}}^{h}F_{2,k}(y)dyd\overline{z}.
\end{eqnarray*}}
\noindent
We determine similarly an expansion of the fluid pressure to any fixed order.

\subsection{Lubrication theory}

The fluid velocities and pressures are
expanded with respect to $\alpha,\beta,\delta\ll 1$, $h_i$ and their space and time derivatives: we use zeroth and first
order expansions of $u_i$ to obtain respectively inviscid and viscous 
conservation laws for $h_1,h_2$. Then we study the spectral stability of steady states.

\subsubsection{Inviscid and viscous conservation laws}

We first compute an inviscid system of conservation laws, which is the
analogous of  Burgers equations in the case of a single fluid
layer. The fluids velocities are given by $\displaystyle
u_i=u_i^{(0)}+\mathcal{O}(\alpha+\beta+\delta)$. The discharge rates
$q_i, i=1,2$ are expanded as
\begin{equation}\label{exp_qi}
\begin{array}{ll}
\displaystyle
q_1=\int_0^{h_1}u_1=\lambda h_1^2\Big(\frac{\rho h_2}{2}+\frac{h_1}{3}\Big)+\mathcal{O}(\alpha+\beta+\delta),\\
\displaystyle
q_2=\int_{h_1}^h u_2=\lambda h_1h_2(\rho h_2+\frac{h_1}{2})+\frac{\lambda\rho}{\nu}\frac{h_2^3}{3}.+\mathcal{O}(\alpha+\beta+\delta).
\end{array}
\end{equation}
\noindent
Inserting (\ref{exp_qi}) into the mass conservation laws
(\ref{cons_mass0}) yields
\begin{equation}\label{burg_bivis}
\begin{array}{ll}
\displaystyle
\partial_t h_1+\partial_x\Big(\lambda h_1^2\big(\frac{\rho h_2}{2}+\frac{h_1}{3}\big)\Big)=\mathcal{O}(\alpha+\beta+\delta),\\
\displaystyle
\partial_t h_2+\partial_x\Big(\lambda h_1h_2\big(\rho h_2+\frac{h_1}{2}\big)+\frac{\lambda\rho}{\nu}\frac{h_2^3}{3}\Big)=\mathcal{O}(\alpha+\beta+\delta).
\end{array}
\end{equation}
\noindent
We drop $\mathcal{O}(\alpha+\beta+\delta)$ terms in (\ref{burg_bivis})
and obtain a system of partial differential equations for $(h_1,h_2)$ 
in a closed form. A necessary condition of stability of steady state
is that (\ref{burg_bivis}) is a {\it hyperbolic system}. When this
condition is satisfied, we obtain a useful information on the
group velocities $\Lambda_+>\Lambda_-$ of low frequency perturbations
(respectively the velocities at the free surface and at the fluid interface): 
{\setlength\arraycolsep{1pt}
\begin{eqnarray}
\displaystyle
\Lambda_{\pm}&=&\frac{\lambda}{4\nu}\big(2\rho h_2^2+6\rho\nu h_1 h_2+3\nu
h_1^2\pm\sqrt{\Delta}\big)\nonumber\\
\displaystyle
\Delta&=&\left(h_1^2\left(2\rho h_2+h_1\right)+8\rho^2h_1^2h_2^2\right)\nu^2+4\rho h_1h_2^2\left(2\rho h_2-h_1\right)\nu+4\rho^2h_2^4\nonumber
\end{eqnarray}}
\noindent 
Strict hyperbolicity is ensured if and only if $\Delta>0$. As it is a quadratic form in $\nu$ the discriminant of $\Delta$ is $-128\rho^3h_1^2h_2^5\left(h_1+\rho h_2\right)$, which is negative when $h_1$, $h_2$, $\rho$ are strictly positive. Then $h_1>0$ and $h_2>0$ ensure that the system (\ref{burg_bivis}) is strictly hyperbolic.

\noindent
Next, we use first order expansions of fluid velocities 
$$
\displaystyle
u_i=u_i^{(0)}+u_i^{(1)}+\mathcal{O}\big((\alpha+\beta+\delta)^2\big)
$$
\noindent
to determine a more accurate system of equations. Inserting the
expansion of $q_i, i=1,2$ into the mass
conservation laws yields a system of Benney's equations (or
Kuramoto-Sivashinsky when surface tension is considered)  
{\setlength\arraycolsep{1pt}
\begin{eqnarray}\label{Benn_Sys}
\displaystyle
\partial_t\left(\begin{array}{c} h_1\\h_2\end{array}\right)+\partial_x\left(\begin{array}{c}\displaystyle\lambda h_1^2(\frac{h_1}{3}+\frac{\rho h_2}{2})\\
                                                                                                                                        \displaystyle \lambda h_1h_2(\rho h_2+\frac{h_1}{2})+\frac{\lambda\rho}{3\nu}h_2^3
                                                                                                               \end{array}\right)&=&\lambda\beta\partial_x\Big(d(h_i)\partial_x\left(\begin{array}{c}h_1\\h_2\end{array}\right)\Big)\nonumber\\&+&\beta\partial_x\Big(K(h_i)\partial_x^3\left(\begin{array}{c}h_1\\h_2\end{array}\right)\Big),
\end{eqnarray}}
\noindent
with the viscous coefficients $d_{i,j}$ defined as $\displaystyle d_{i,j}=\frac{\cot\theta}{R_e}d_{i,j,1}-\lambda\,d_{i,j,2}$
{\setlength\arraycolsep{1pt}
\begin{eqnarray}\label{def_dij}
\displaystyle
d_{1,1,1}&=&h_1^2(\frac{h_1}{3}+\frac{\rho h_2}{2}),\quad d_{1,2,1}=\rho h_1^2(\frac{h_1}{3}+\frac{h_2}{2}),\nonumber\\
\displaystyle
d_{2,1,1}&=&\frac{\rho h_2^3}{3\nu}+\rho h_1h_2^2+h_2\frac{h_1^2}{2},\quad d_{2,2,1}=\frac{\rho h_2^3}{3\nu}+\rho h_1h_2^2+\rho h_2\frac{h_1^2}{2},\nonumber\\
\displaystyle
d_{1,1,2}&=&\frac{h_1^2}{\nu}\Big(\frac{2h_1^4}{15}+\frac{71\rho\nu}{120}h_1^3h_2+\frac{23\rho^2\nu}{24}h_1^2h_2^2+(\frac{\rho^2}{6}+\frac{\rho^2\nu}{2})h_1h_2^3+\frac{\rho^3}{6}h_2^4\Big),\nonumber\\
\displaystyle
d_{1,2,2}&=&\frac{\rho h_1^2}{\nu^2}\Big(\frac{2\nu^2 h_1^4}{15}+\frac{71\rho\nu^2}{120}h_1^3h_2+(\frac{5\rho\nu}{24}+\frac{3\rho^2\nu^2}{4})h_1^2h_2^2+\frac{2\rho^2\nu}{3}h_1h_2^3+\frac{\rho^2}{6}h_2^4\Big),\nonumber\\
\displaystyle
d_{2,1,2}&=&\frac{h_2}{\nu^2}\Big(\frac{5\nu^2h_1^5}{24}+\frac{25\rho\nu^2}{24}h_1^4h_2+(\frac{11\rho^2\nu^2}{6}+\frac{\rho\nu}{6})h_1^3h_2^2\nonumber\\
\displaystyle
&&+(\rho^3\nu^2+\frac{5\rho^2\nu}{6})h_1^2h_2^3+(\frac{2\rho^2}{15}+\frac{2\rho^3\nu}{3})h_1h_2^4+\frac{2\rho^3}{15}h_2^5\Big),\nonumber\\
\displaystyle
d_{2,2,2}&=&\frac{\rho
  h_2}{\nu^2}\Big(\frac{5\nu^3}{24}h_1^5+\frac{25\rho\nu^3}{24}h_1^4h_2+(\frac{3\rho^2\nu^3}{2}+\frac{\rho\nu^2}{2})h_1^3h_2^2\nonumber\\
\displaystyle
&&+\frac{11\rho^2\nu^2}{6}h_1^2h_2^3+\frac{4\rho^2\nu h_1h_2^4}{5}+\frac{2\rho^2}{15}h_2^5\Big).
\end{eqnarray}}
\noindent
The surface tension terms are given by
\begin{equation}
\begin{array}{lll}
\displaystyle
K_{1,1}=h_1^2\big((\kappa_1+\kappa_2)\frac{h_1}{3}+\kappa_2\frac{h_2}{2}\big),
\quad K_{1,2}=\kappa_2\,h_1^2\big(\frac{h_1}{3}+\frac{h_2}{2}\big),\\
\displaystyle
K_{2,1}=\frac{\kappa_2h_2^3}{3\nu}+\kappa_2h_2^2h_1+(\kappa_1+\kappa_2)\frac{h_2h_1^2}{2},\\
\displaystyle
K_{2,2}=\frac{\kappa_2 h_2^3}{3\nu}+\kappa_2h_1h_2^2+\frac{\kappa_2h_2h_1^2}{2}.
\end{array}
\end{equation}

\noindent
This system is in agreement with the one in \cite{kliak}. In the low
frequency regime, the system (\ref{Benn_Sys}) of viscous conservation
laws provides a criterion of spectral stability for the steady
solutions which is consistent with the one given by Orr-Sommerfeld
equations (if $R_e, F=O(1)$). One goal of this paper is to study the
formation of roll-waves in bi-layer flows. In the single layer case,
they are the result ofthe competition between buoyancy and
inertia. Therefore, we focus on the competition between inertia and
buoyancy and their interaction with convective terms to describe the
onset of roll-waves.

\subsubsection{Spectral Stability of Steady States}

Let us linearise (\ref{Benn_Sys}) at a constant state $(\overline{h}_1,\overline{h}_2)$: 
\begin{equation}\label{lin_benn}
\displaystyle
\partial_t\left(\begin{array}{c}h_1\\h_2\end{array}\right)+J(\overline{h}_i)\partial_x\left(\begin{array}{c}h_1\\h_2\end{array}\right)=\lambda\beta d(\overline{h}_i)\partial_{xx}\left(\begin{array}{c}h_1\\h_2\end{array}\right). 
\end{equation}
\noindent
Without loss of generality, we assume  $\lambda\beta=1$. We have neglected the
contribution of surface tension as they are not relevant in the low
frequency regime. The dispersion relation is given by
\begin{equation}\label{disp0}
\displaystyle
{\rm det}(\Lambda Id+ik\,J(h)+k^2\,d(h))=0,\quad \forall\,k\in\mathbb{R}.
\end{equation}
\noindent
and we assume $|k|\ll 1$. We expand $\Lambda_j, j=1,2$ as
$\Lambda_j=ik\tilde{\Lambda}_j$. Equation (\ref{disp0}) then reads
\begin{equation*}
\displaystyle
{\rm det}\big(\tilde{\Lambda}+J(\overline{h})\big)=ik{\rm tr}\Big({\rm com}\big(\tilde{\Lambda} Id+J(\overline{h})\big)^Td(\overline{h})\Big)+\mathcal{O}(k^2). 
\end{equation*}
\noindent
System (\ref{burg_bivis}) is strictly hyperbolic : the eigenvalues $\overline{\Lambda}_i$ of $J(\overline{h})$ are real and $\overline{\Lambda}_1>\overline{\Lambda}_2$. Then $\Lambda_1(k),\Lambda_2(k)$ expand at $k=0$ as 
\begin{equation}
\displaystyle
\Lambda_j(k)=-ik\overline{\Lambda}_j-k^2\frac{{\rm tr}\Big({\rm com}\big(J(\overline{h}_i)-\overline{\Lambda}_j Id\big)^Td(\overline{h}_i)\Big)}{{\rm tr} \big(J(\overline{h}_i)\big)-2\overline{\Lambda}_j}+\mathcal{O}(k^3).
\end{equation}
 \noindent
 As a result, stationary solutions are stable if
 \begin{equation}\label{st_vis}
 \displaystyle
 {\rm tr}\Big({\rm com}\big(J(\overline{h})-\overline{\Lambda}_1 Id\big)^Td(\overline{h})\Big)<0,\quad {\rm tr}\Big({\rm com}\big(J(\overline{h})-\overline{\Lambda}_2 Id\big)^Td(\overline{h})\Big)>0.
 \end{equation}

\noindent
We consider two particular situations: stable and unstable density 
stratification: we will see that in some particular
situations, the Rayleigh Taylor instability may be suppressed by the convection.
We also study the influence of inertia on stability properties.\\

\noindent
The spectral stability conditions (\ref{st_vis}) have the simple form  
$$
\displaystyle
a_1(\rho)\frac{{\rm cotan}\theta}{R_e}<\lambda\,a_2(\rho),\quad b_1(\rho)\frac{{\rm cotan}\theta}{R_e}>\lambda\,b_2(\rho).
$$
\noindent
It is easily seen that $a_i<0$: the free surface is then stable if 
$$
\displaystyle 
R_e\leq \frac{a_1(\rho)}{\lambda a_2(\rho)}{\rm cotan}\theta.
$$
\noindent
The situation is more involved for the fluid interface where $b_i$ can
change sign. If $b_1(\rho)b_2(\rho)<0$, the interface is stable when
$b_1(\rho)>0$ and unstable otherwise. If $b_1(\rho)b_2(\rho)>0$, the
fluid interface is stable if
$$
\displaystyle b_1(\rho)\big(R_e-\frac{b_1(\rho)}{\lambda b_2(\rho)}{\rm
  cotan}\theta\big)<0. 
$$ 

\noindent
In order to simplify the spectral analysis, we set
$\overline{h}_i=1, i=1,2$ and consider the
cases $\nu<1$, $\nu=1$ and $\nu>1$ (stratification
in viscosity). We have determined stability curves
$R_e=f_k(\rho){\rm cotan}\theta, k=1,2$ associated to the surface mode
$k=1$ and the interfacial mode $k=2$.\\

\mathversion{bold}{\bf Case 1: $\nu<1$.}\mathversion{normal} We have chosen here $\nu=0.3$, $\nu=0.7$ and
$\nu=0.9$ that gives a good representation of all possible scenarii
that arises as $\rho$ varies. Let us first check the case
$\nu=0.3$. We have represented critical curves in picture
\ref{fig1}.
\begin{figure}[h!]
\begin{center}
\begin{minipage}[t]{0.45\linewidth}
\begin{center}
\fbox{\includegraphics[width=\linewidth]{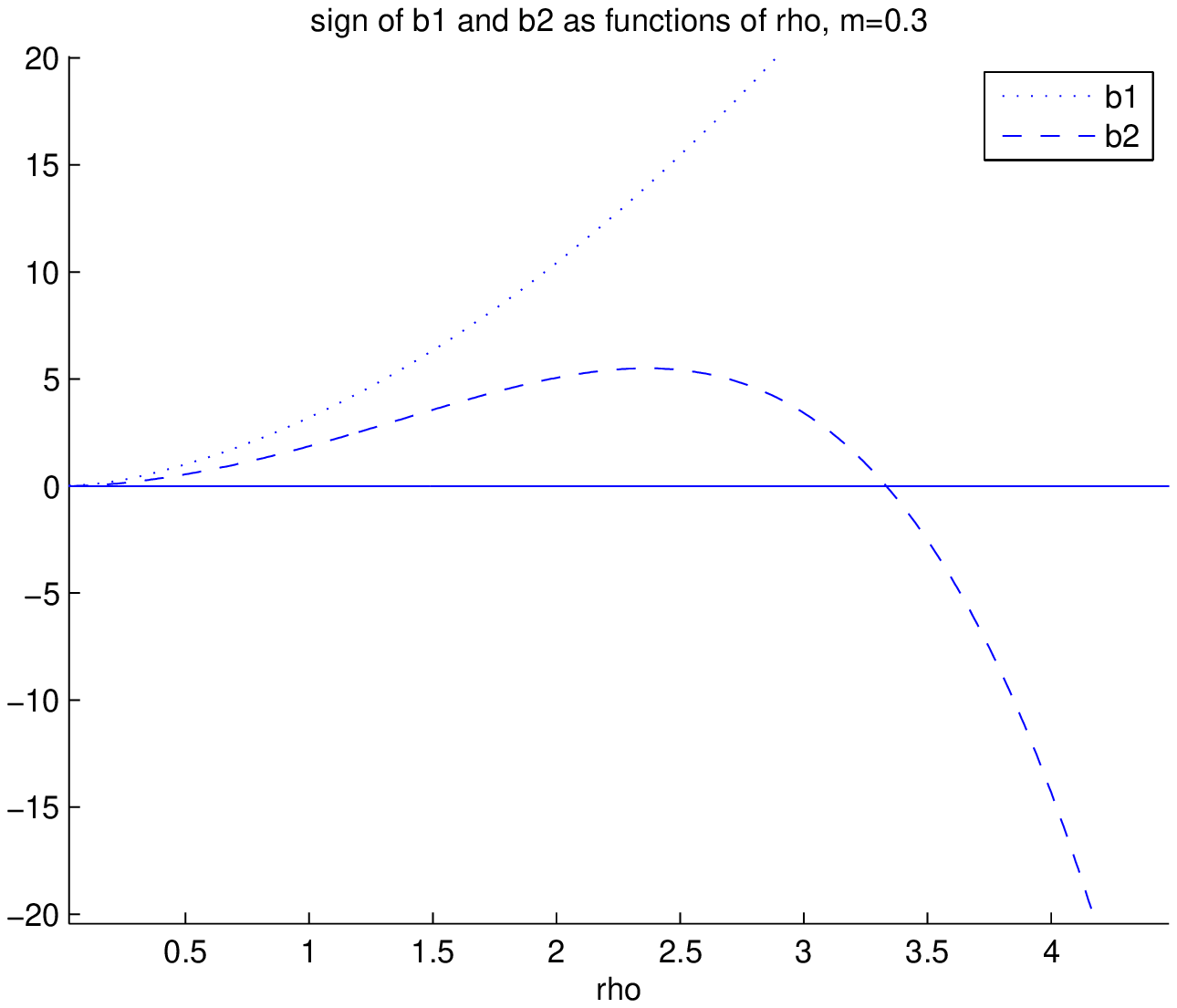}}
\end{center}
\end{minipage}
\hfill
\begin{minipage}[t]{0.45\linewidth}
\begin{center}
\fbox{\includegraphics[width=\linewidth]{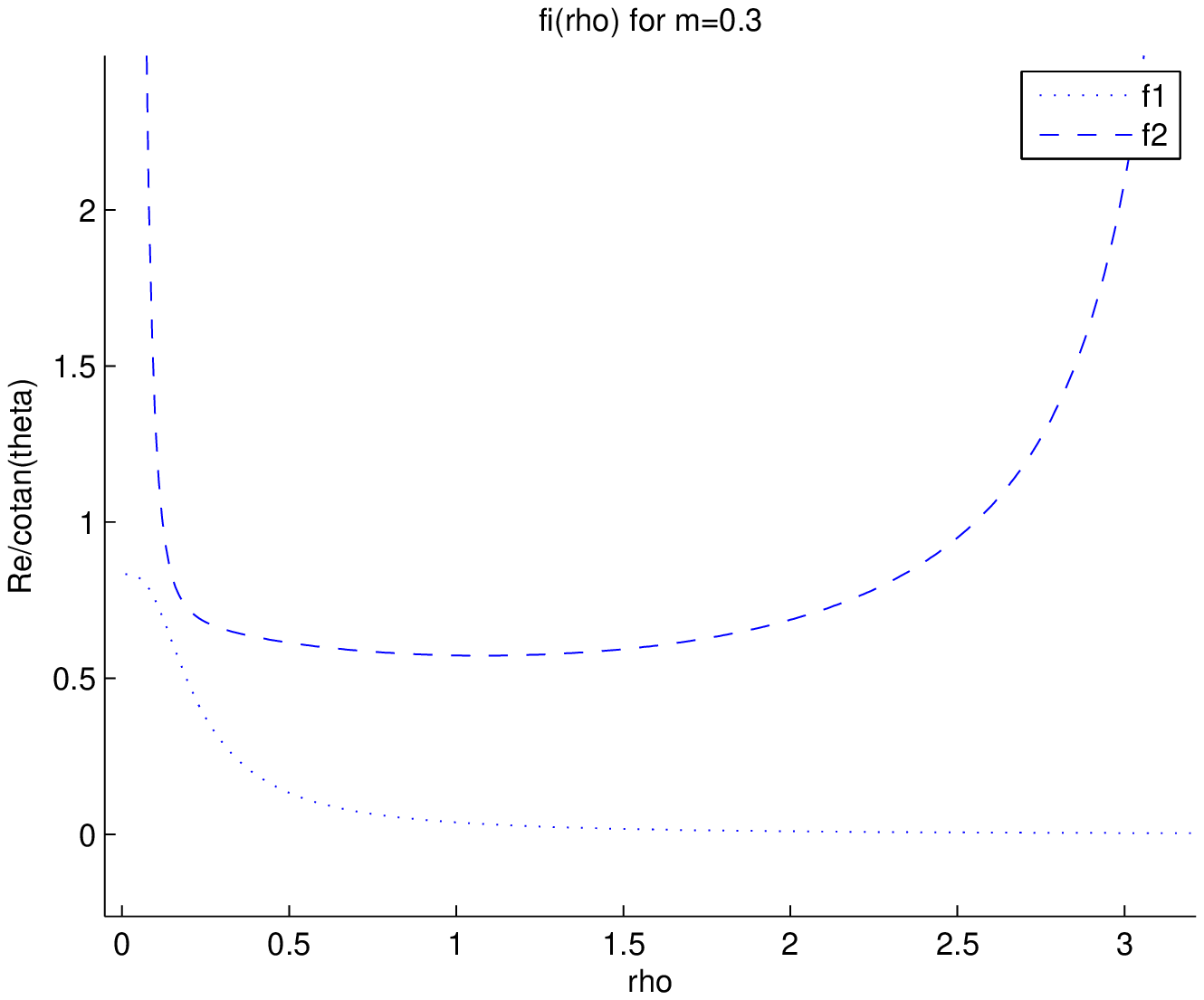}}
\end{center}
\end{minipage}
\end{center}
\caption{\label{fig1}Values of $b_i,i=1,2$ and the critical curves $\displaystyle
  f_i(\rho)=\frac{R_e}{{\rm cotan}\theta}$ for $\nu=0.3$} 
\end{figure}
There exists $\rho_c\approx 3.3$ above which $b_2<0$ otherwise both
$b_i>0, i=1,2$. Then, for $\rho>\rho_c$, the
interfacial mode is stable whereas the full system is spectrally
stable if $\R_e<f_1(\rho){\rm cotan}\theta$. If $\rho<\rho_c$,
$f_1(\rho)<f_2(\rho)$: if $R_e$ is sufficiently small, the flow is
stable and as $R_e$ is increased, the surface mode is
destabilized before the interfacial mode.

\begin{figure}[h!]
\begin{center}
\begin{minipage}[t]{0.45\linewidth}
\begin{center}
\fbox{\includegraphics[width=\linewidth]{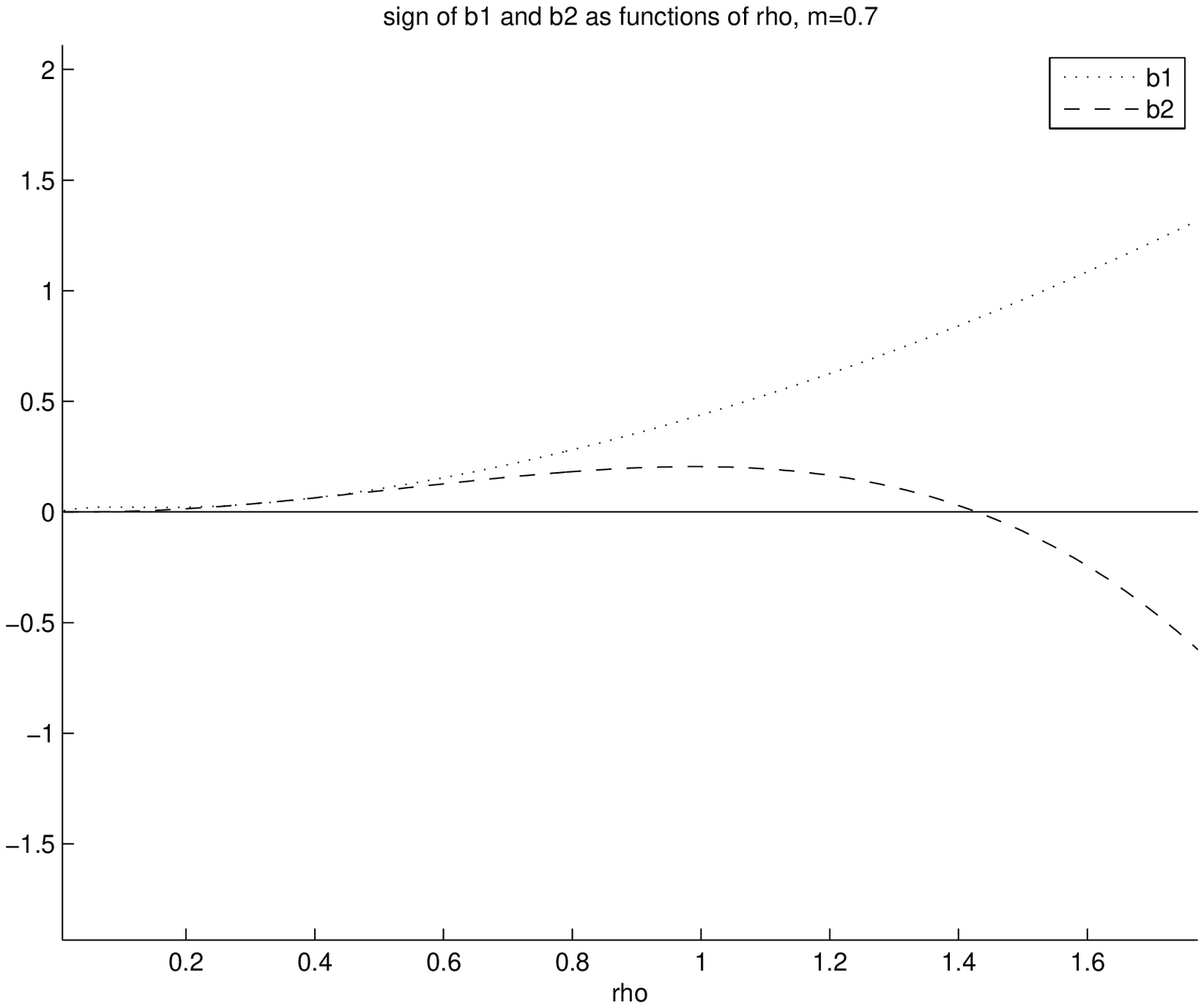}}
\end{center}
\end{minipage}%
\hfill
\begin{minipage}[t]{0.45\linewidth}
\begin{center}
\fbox{\includegraphics[width=\linewidth]{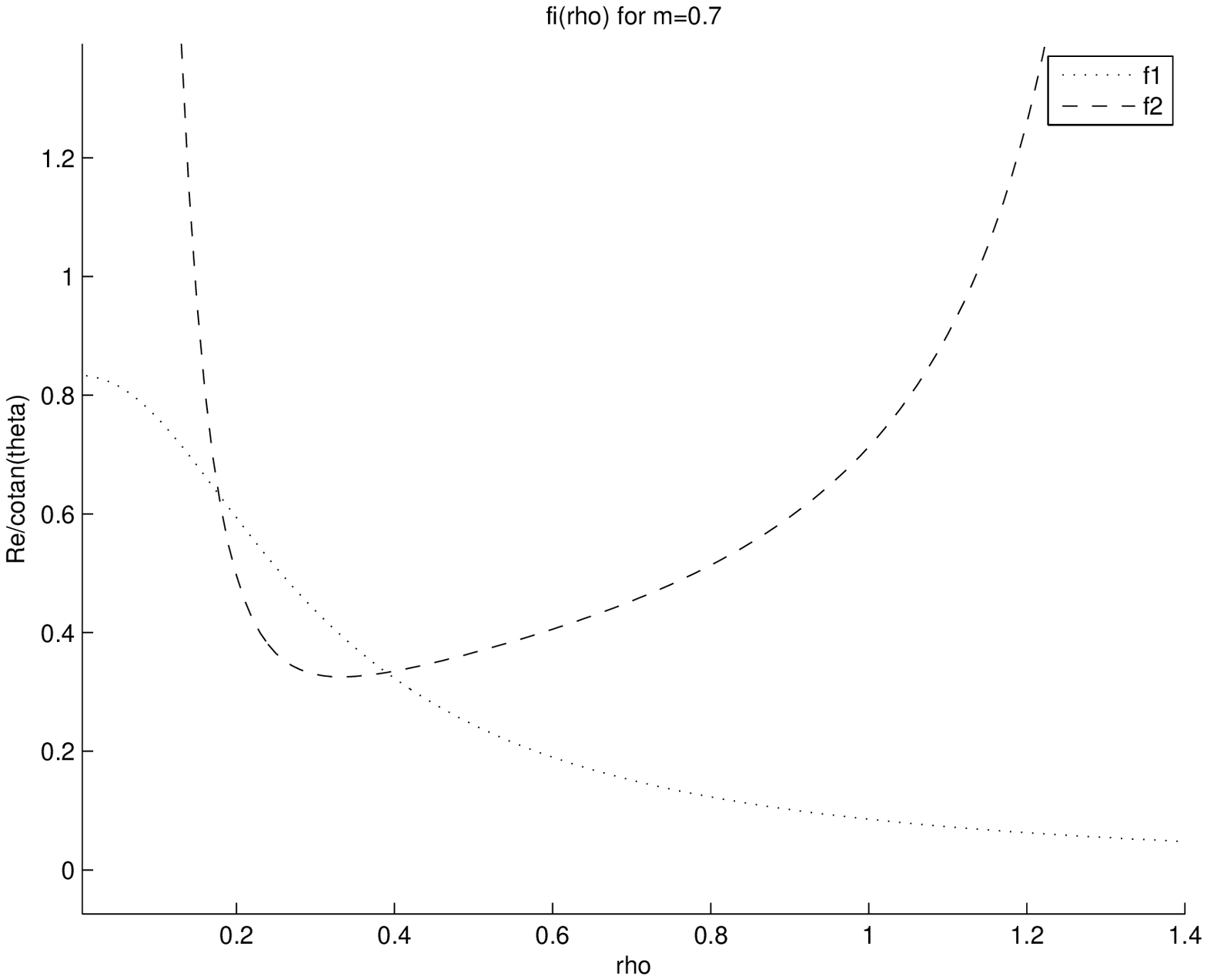}}
\end{center}
\end{minipage}
\end{center}
\caption{\label{fig2}Values of $b_i,i=1,2$ and the critical curves $\displaystyle
  f_i(\rho)=\frac{R_e}{{\rm cotan}\theta}$ for $\nu=0.7$} 
\end{figure}

\begin{figure}[h!]
\begin{center}
\begin{minipage}[t]{0.45\linewidth}
\begin{center}
\fbox{\includegraphics[width=\linewidth]{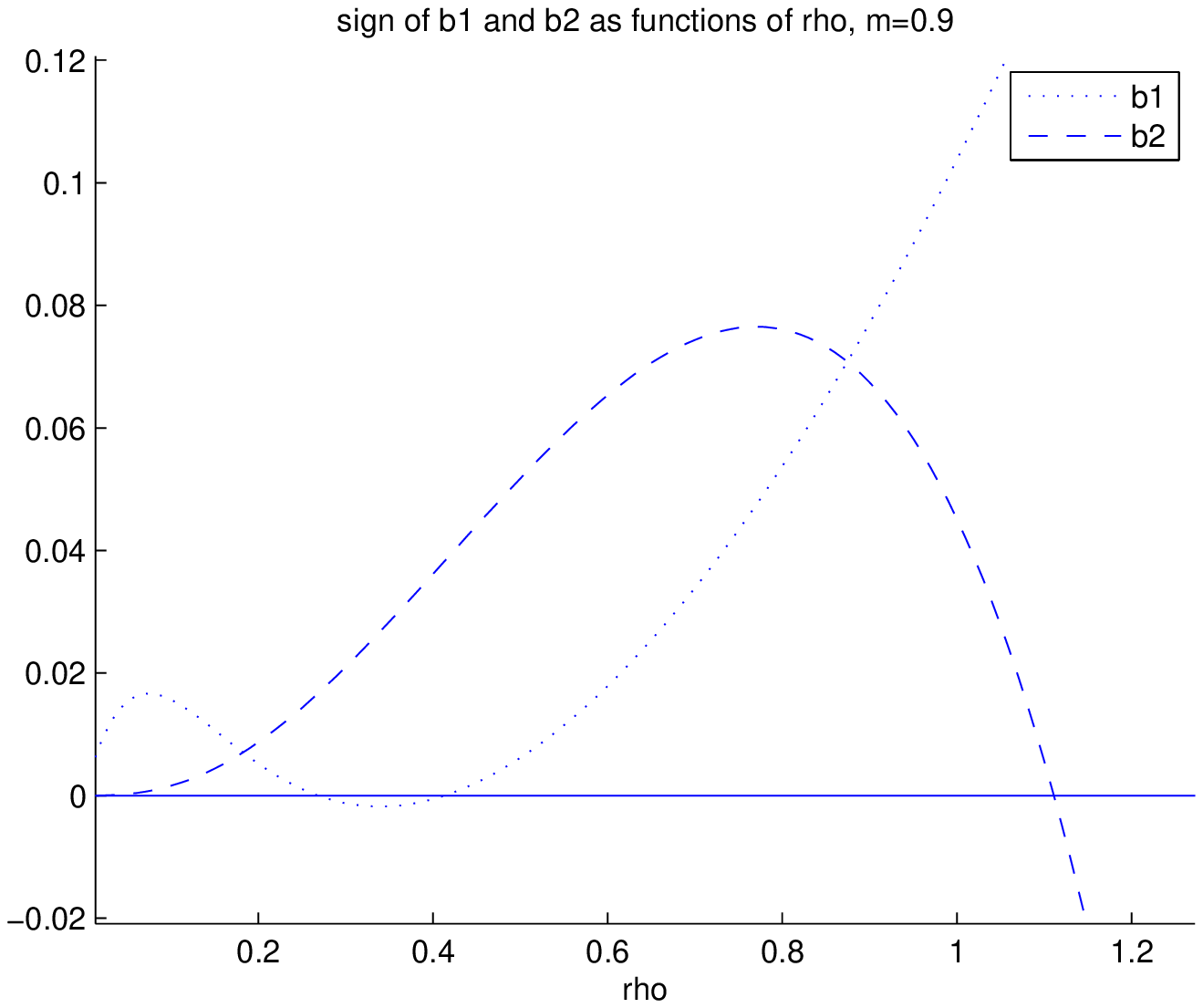}}
\end{center}
\end{minipage}%
\hfill
\begin{minipage}[t]{0.45\linewidth}
\begin{center}
\fbox{\includegraphics[width=\linewidth]{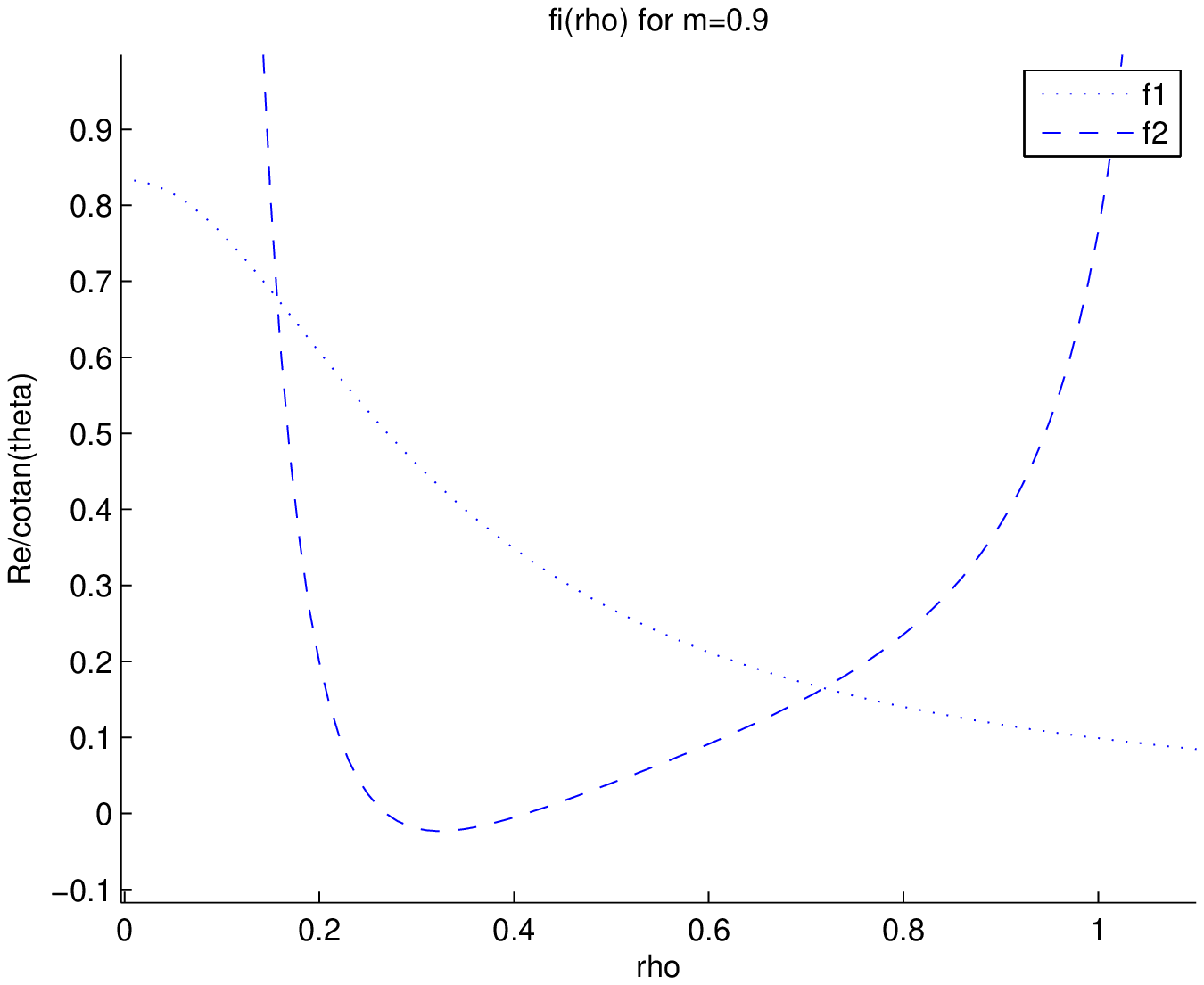}}
\end{center}
\end{minipage}
\end{center}
\caption{\label{fig3}Values of $b_i,i=1,2$ and the critical curves $\displaystyle
  f_i(\rho)=\frac{R_e}{{\rm cotan}\theta}$ for $\nu=0.9$} 
\end{figure}

Next we consider the case $\nu=0.7$. The critical curves are represented in
picture \ref{fig2}.
There exists $\rho_c\approx 1.4$ above which the
interfacial mode is always unstable whereas the surface mode is stable
for $R_e<f_2(\rho){\rm cotan}\theta$. If $\rho<\rho_c$, there exists
$\rho_1<\rho_2$ so that for any $\rho<\rho_1$ or $\rho_2<\rho<\rho_c$,
the scenario is identical to the previous case: as $R_e$ is increased,
the surface mode is destabilized before the interface
mode. If $\rho_1<\rho<\rho_2$, the interfacial mode is 
destabilized first as $R_e$ is increased. The case $\nu=0.9$ is
similar, except that for $\rho_1<\rho<\rho_2$, the interfacial
mode is always unstable (see picture \ref{fig3}).
\\

\begin{figure}[ht]
\begin{center}
\begin{minipage}[t]{0.45\linewidth}
\begin{center}
\fbox{\includegraphics[width=\linewidth]{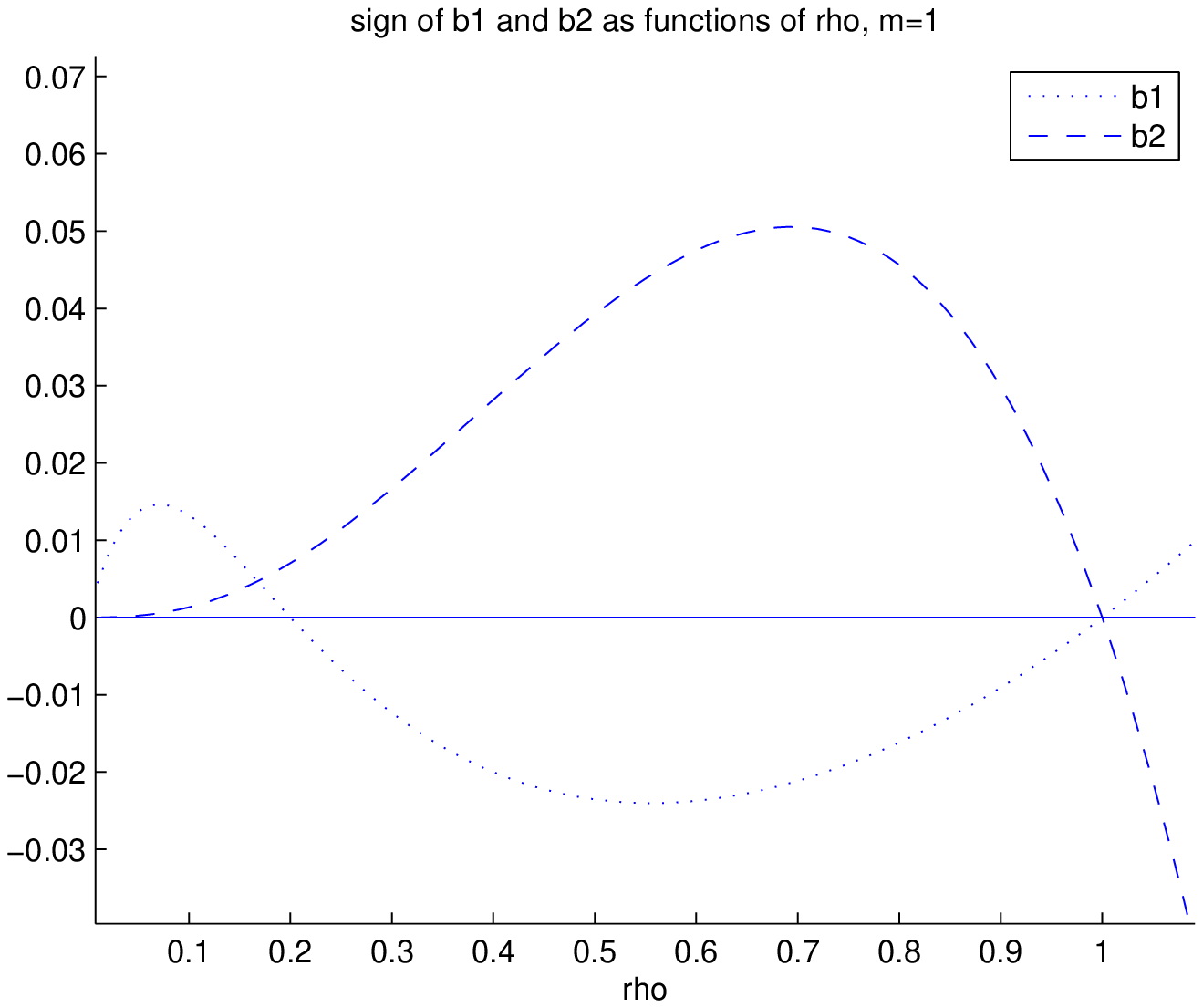}}
\end{center}
\end{minipage}%
\hfill
\begin{minipage}[t]{0.45\linewidth}
\begin{center}
\fbox{\includegraphics[width=\linewidth]{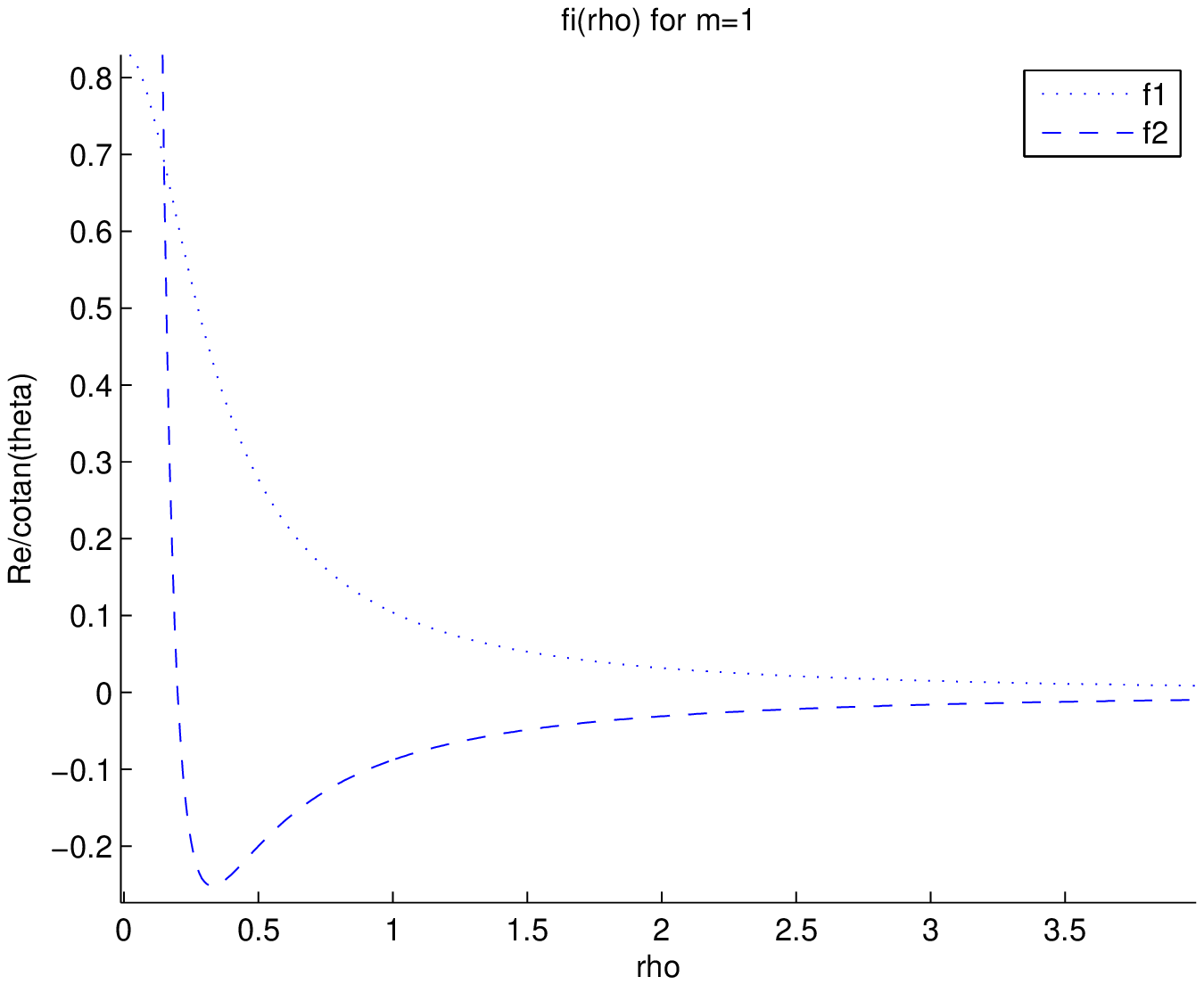}}
\end{center}
\end{minipage}
\end{center}
\caption{\label{fig4}Values of $b_i,i=1,2$ and the critical curves $\displaystyle
  f_i(\rho)=\frac{R_e}{{\rm cotan}\theta}$ for $\nu=1$} 
\end{figure}

\mathversion{bold}
{\bf Case 2: $\nu=1$.}\mathversion{normal} There exists $\rho_c$ above which
the interfacial mode is always stable. If $\rho<\rho_c$, there exists
$\rho_1<\rho_c$ so that the interface mode is always unstable if
$\rho_1<\rho<\rho_c$. If $\rho<\rho_1$, the surface mode is 
destablized first $R_e$ is increased (see figure \ref{fig4}).\\

\begin{figure}[ht]
\begin{center}
\begin{minipage}[t]{0.45\linewidth}
\begin{center}
\fbox{\includegraphics[width=\linewidth]{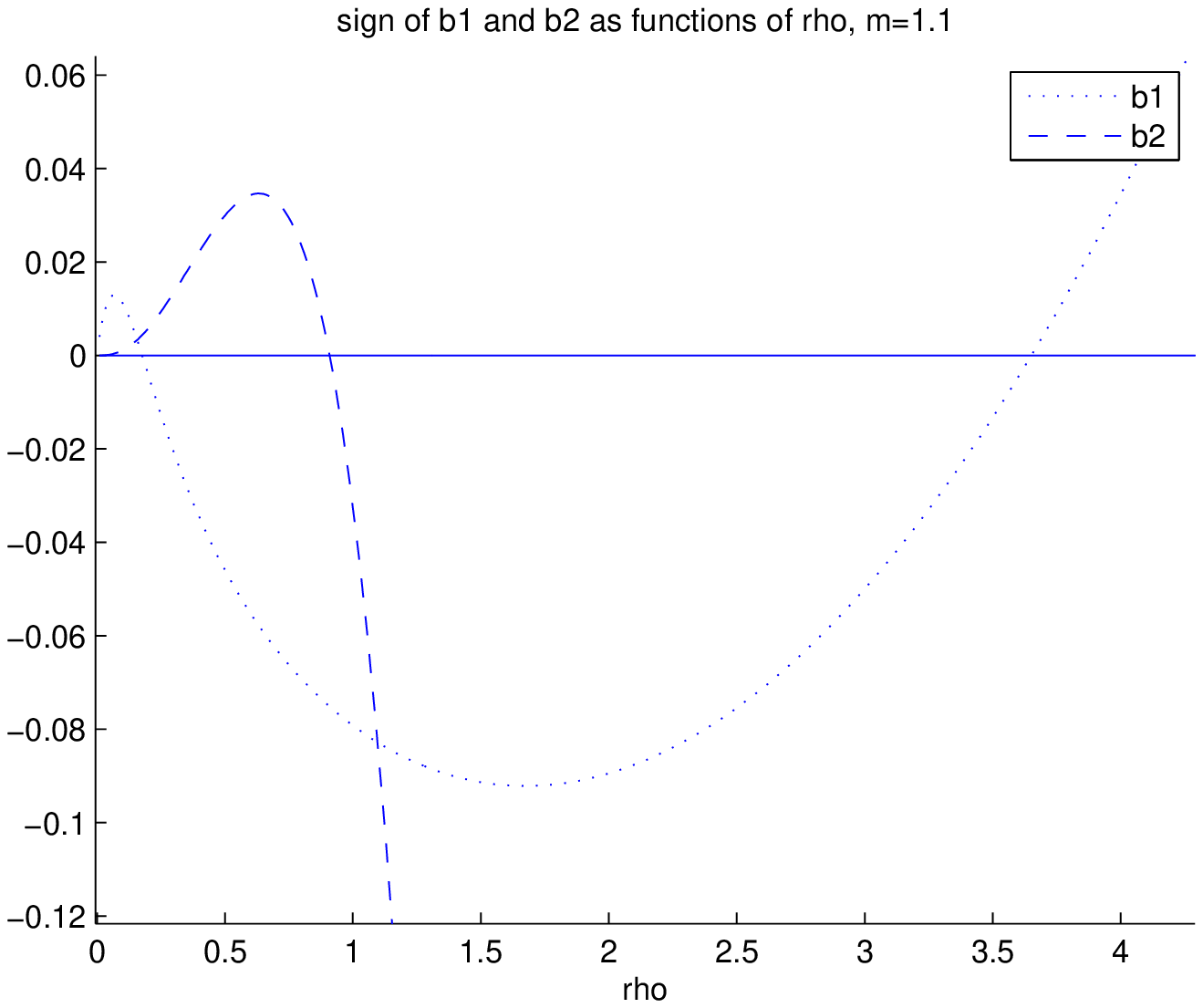}}
\end{center}
\end{minipage}%
\hfill
\begin{minipage}[t]{0.45\linewidth}
\begin{center}
\fbox{\includegraphics[width=\linewidth]{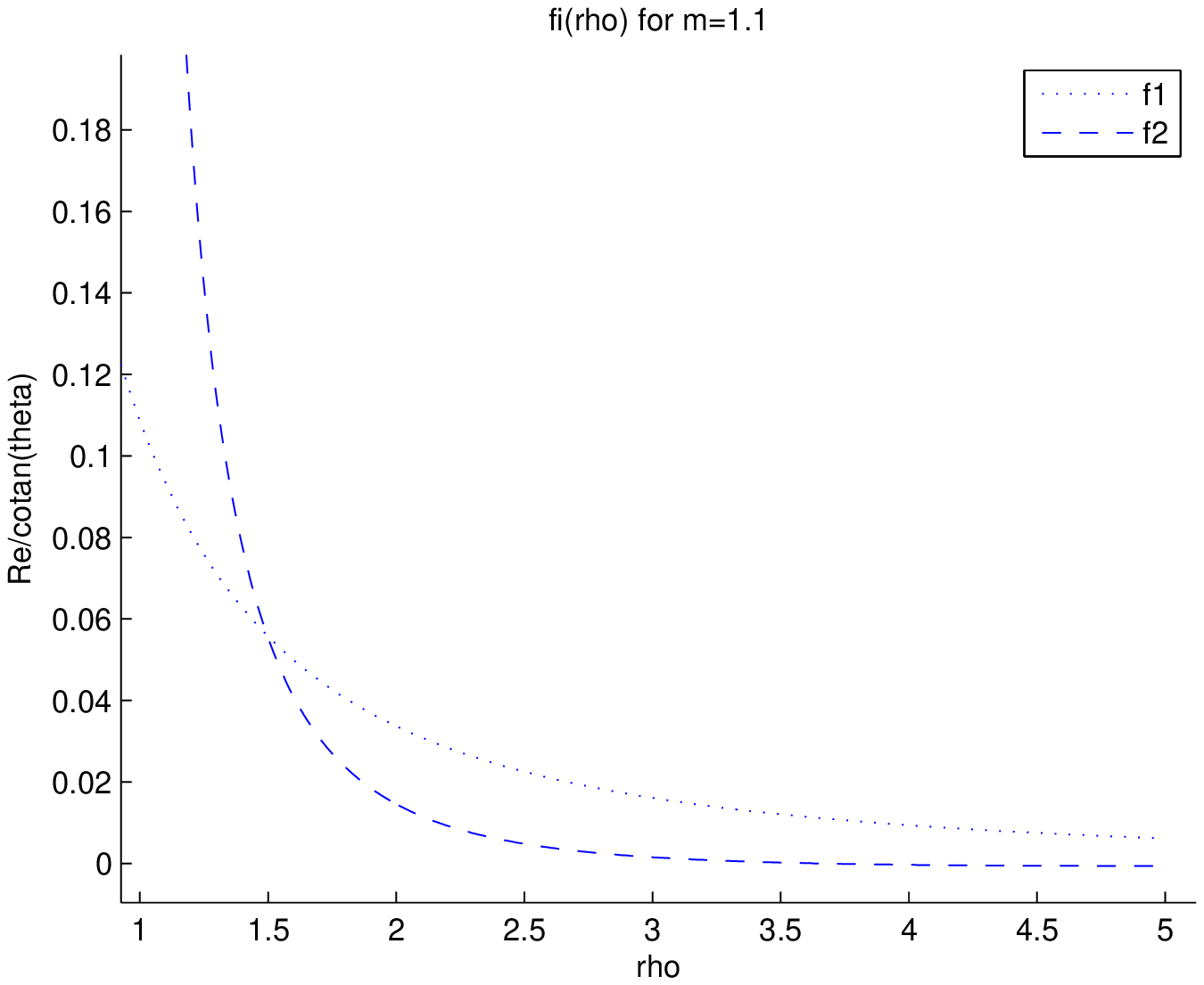}}
\end{center}
\end{minipage}
\end{center}
\caption{\label{fig5}Values of $b_i,i=1,2$ and the critical curves $\displaystyle
  f_i(\rho)=\frac{R_e}{{\rm cotan}\theta}$ for $\nu=1.1$} 
\end{figure}

\begin{figure}[ht]
\begin{center}
\begin{minipage}[t]{0.45\linewidth}
\begin{center}
\fbox{\includegraphics[width=\linewidth]{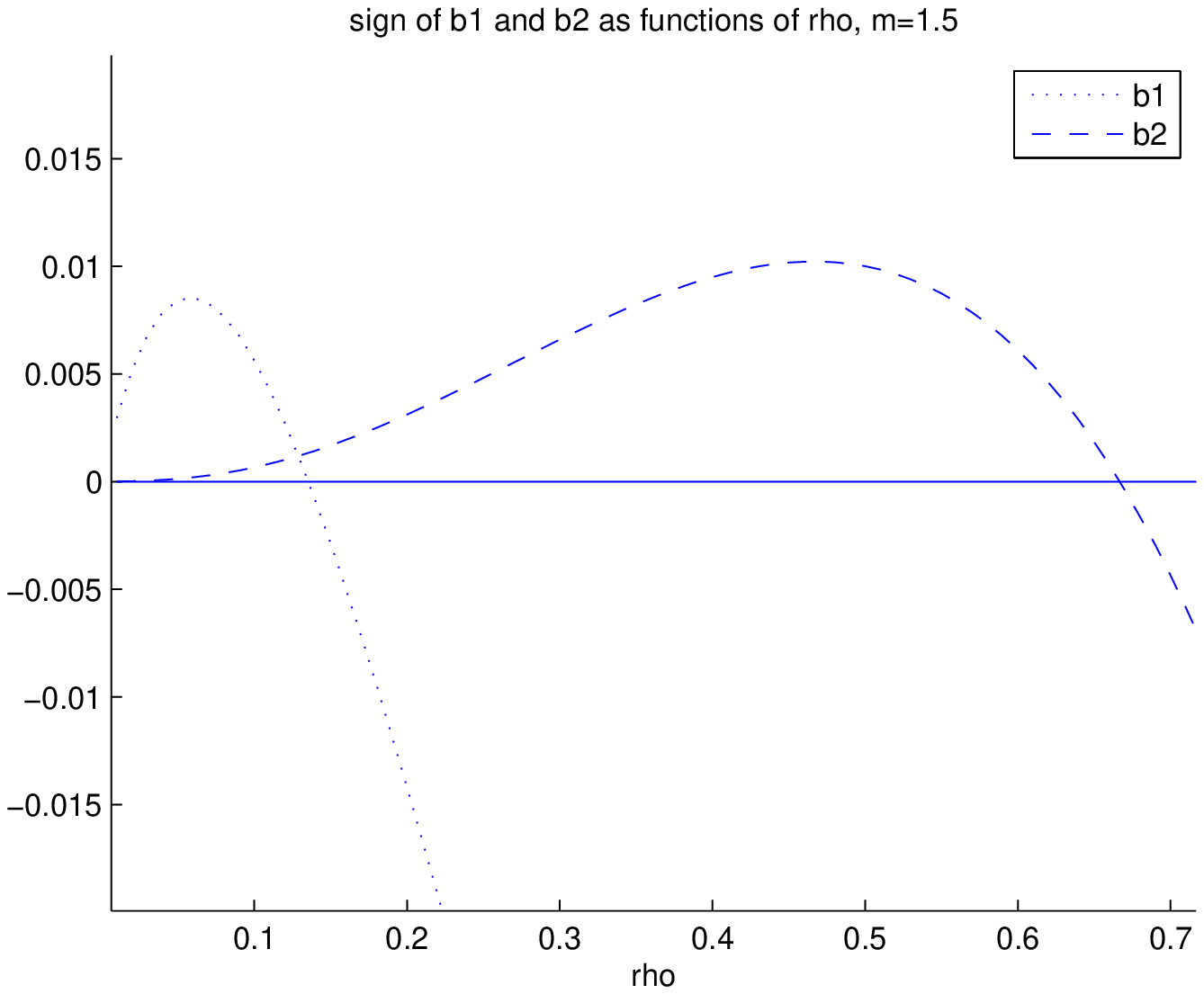}}
\end{center}
\end{minipage}%
\hfill
\begin{minipage}[t]{0.45\linewidth}
\begin{center}
\fbox{\includegraphics[width=\linewidth]{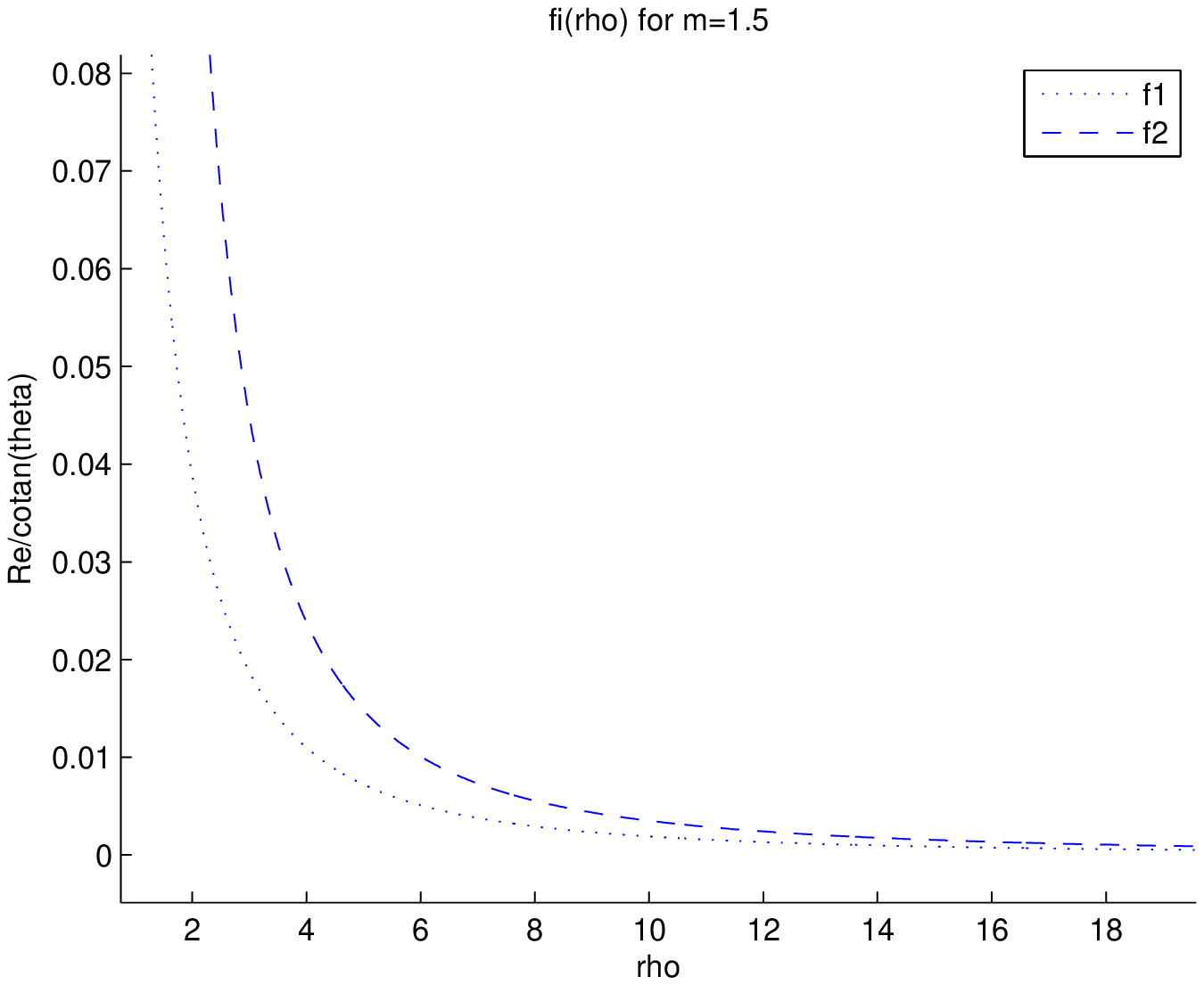}}
\end{center}
\end{minipage}
\end{center}
\caption{\label{fig6}Values of $b_i,i=1,2$ and the critical curves $\displaystyle
  f_i(\rho)=\frac{R_e}{{\rm cotan}\theta}$ for $\nu=1.5$} 
\end{figure}

\mathversion{bold}
{\bf Case 3: $\nu>1$.} \mathversion{normal} We have chosen $\nu=1.1$ and
$\nu=1.5$, which gives a good representation of all possible scenarii
that arises as $\rho$ varies. We first consider $\nu=1.1$
(see figure \ref{fig5}).
There exists $\rho_c\approx 3.5$ above which the interfacial mode is
always stable. Assume $\rho<\rho_c$ and denote
$\rho_1<\rho_2$ respectively the first zeros of $b_1$ and the zero of
$b_2$. If $\rho_1<\rho<\rho_2$, the interfacial mode is always
unstable and if $\rho<\rho_1$, the situation is similar to previous
cases when $\rho$ is small. If $\rho>\rho_2$, the interfacial mode is 
stable only if $R_e>f_1(\rho){\rm cotan}\theta$. As a consequence, even at low
Reynolds number, the interfacial mode is unstable. It is easily seen 
that there exists $\rho_2<\rho_3<\rho_c$ such that the flow is always 
unstable if $\rho_2<\rho<\rho_3$, stable if
$\rho_3<\rho<\rho_3$ and 
$$
\displaystyle
f_2(\rho){\rm cotan}\theta<R_e<f_1(\rho){\rm cotan}\theta.
$$ 
Finally, we consider the case $\nu=1.5$ (see figure \ref{fig6}).
If $\rho_1$ is the zero of $b_1$, then for any $\rho>\rho_1$,
the flow is always unstable under long wavelength perturbations. If
$\rho<\rho_1$, the situation is similar to the cases $\nu<1$ when
$\rho$ is stable: the flow is stable at low Reynolds number and the
flow destabilizes first through the surface mode or the interfacial mode. 

\subsection{Shallow water theory}

The viscous conservation laws which govern the evolution of $h_i$ are
sufficient to obtain a consistent stability criterion of constant
states in the low frequency regime. However, the solutions to this
system blow up in finite time when the flow is unstable and may lead
to some inaccuracy in the description of the motion of bi-layer
flows. In what follows, we consider shallow water models: indeed in
the case of a single fluid layer, they sustain
nonlinear waves, so called ``roll-waves'' which are well known
hydrodynamic instabilities. As a consequence, shallow water models are
useful to describe the transition to instability in shallow flows. Up
to our knowledge, there's no consistent shallow water model which
describes bi-layer flows down a ramp (the single layer case was
treated only recently \cite{r-quil1},\cite{Vila}).\\

\noindent
Let us keep  $\mathcal{O}(1)$ and $\mathcal{O}(\varepsilon^{-1})$
terms in (\ref{av_eq_mom}):
{\setlength\arraycolsep{1pt}
\begin{eqnarray}\label{av_eq_mom0}
\displaystyle
\partial_t\big(\int_0^{h_1}u_1\big)+\partial_x\Big(\int_0^{h_1}u_1^2+\frac{p_1}{F^2}\Big)&+&\kappa_1\partial_x
  h_1\partial_{xx}h_1=\nonumber\\
\frac{1}{\varepsilon R_e}\big(\lambda h_1&+&\nu\partial_z u_2(h_1)-\partial_z u_1(0)\big)+\frac{p_2(h_1)\partial_x h_1}{F^2}\nonumber\\
\displaystyle
\partial_t\big(\rho\int_{h_1}^{h}u_2\big)+\partial_x\Big(\rho\int_{h_1}^{h}u_2^2+\frac{p_2}{F^2}\Big)&+&\kappa_2\partial_x
  h\partial_{xx}h=\nonumber\\
\displaystyle  
\frac{1}{\varepsilon R_e}\big(\lambda\rho h_2&-&\nu\partial_z u_2(0)\big)-\frac{p_2(h_1)\partial_x h_1}{F^2}.
\end{eqnarray}}
\noindent
We first compute an expansion of the integrals: the integrals of the
pressures are given by
\begin{equation}\label{dl_av_p}
\begin{array}{lll}
\displaystyle
\int_{h_{i-1}}^{h_{i-1}+h_i}p_i=\int_0^{h_1}p_i^{(0)}+\mathcal{O}(\alpha+\beta+\delta),\quad
i=1,2,\\
\displaystyle
\int_0^{h_1}p_1^{(0)}=c\frac{h_1^2}{2}+\rho ch_1h_2-h_1F^2\big((\kappa_2+\kappa_1)\partial_{xx}h_1-\kappa_2\partial_{xx}h_2\big),\\
\displaystyle
\int_{h_1}^{h}p_2^{(0)}=\rho c\frac{h_2^2}{2}-\kappa_2F^2h_2\partial_{xx}h_2.
\end{array}
\end{equation}
\noindent
whereas the integrals of convection terms are given by
{\setlength\arraycolsep{1pt}
\begin{eqnarray}\label{dl_av_u2}
\displaystyle
\int_{h_{i-1}}^{h_{i-1}+h_i}
u_i^2&=&\int_{h_{i-1}}^{h_{i-1}+h_i}(u_i^{(0)})^2+\mathcal{O}(\alpha+\beta+\delta),\quad
i=1,2\nonumber\\
\displaystyle
\int_0^{h_1}(u_1^{(0)})^2&=&\lambda^2h_1^3\Big(\frac{2}{15}h_1^2+\frac{5}{12}\rho h_1h_2+\frac{\rho^2 h_2^3}{3}\Big),\nonumber\\
\displaystyle
\int_{h_1}^{h}(u_2^{(0)})^2&=&\lambda^2\big(\frac{2\rho^2}{15\nu^2}h_2^5+\frac{2\rho}{3\nu}(\rho
h_1h_2+\frac{h_1^2}{2})h_2^3\big)\nonumber\\
&&+\lambda^2\,h_1^2\big(\rho^2h_2^3+\rho h_2^2h_1+\frac{h_2h_1^2}{4})\big).\nonumber
\end{eqnarray}}
\noindent
In both cases, we only keep $O(1)$ terms. In order to write these
quantities as classical convective terms, we introduce $Q_i, i=1,2$ so that
\begin{equation}
\displaystyle
\int_0^{h_1}(u_1^{(0)})^2=h_1\overline{u}_1^2+Q_1(h_1,h_2),\quad \int_{h_1}^{h}(u_2^{(0)})^2=h_2\overline{u}_2^2+Q_2(h_1,h_2).
\end{equation}
\noindent
Here, $Q_i$ depend on $h_i$ and are defined as
$$
\displaystyle
Q_1=\lambda^2\,h_1^3\Big(\frac{h_1^2}{45}+\frac{\rho h_2}{12}(h_1+\rho h_2)\Big),\quad Q_2=\lambda^2\frac{\rho^2\,h_2^5}{45\,\nu^2}.
$$
\noindent
Inserting (\ref{dl_av_p}) and (\ref{dl_av_u2}) into (\ref{av_eq_mom0}), one finds
{\setlength\arraycolsep{1pt}
\begin{eqnarray*}
\displaystyle
\partial_t (h_1\overline{u}_1)+\partial_x\Big(h_1\overline{u}_1^2&+&\frac{c}{F^2}\big(\rho h_1h_2+\frac{h_1^2}{2}\big)+Q_1\Big)-h_1\partial_{xxx}\big(\kappa_2(h_1+h_2)+\kappa_1h_1\big)\nonumber\\
\displaystyle
&=&\frac{\rho c}{F^2}h_2\partial_x h_1+\frac{sh_1}{\varepsilon F^2}+\frac{1}{\varepsilon R_e}\big(\nu\partial_z u_2(h_1)-\partial_z u_1(0)\big),\\
\displaystyle
\rho\Big(\partial_t (h_2\overline{u}_2)+\partial_x\big(h_2\overline{u}_2^2&+&\frac{c}{F^2}\frac{h_2^2}{2}+Q_2\big)\Big)-\kappa_2h_2\partial_{xxx}h\nonumber\\
\displaystyle
&=&-\frac{\rho c}{F^2}h_2\partial_x h_1+\frac{\rho s\,h_2}{\varepsilon F^2}-\frac{\nu}{\varepsilon R_e}\partial_z u_2(h_1).
\end{eqnarray*}}
\noindent
This system is almost in a closed form. Let us now write $\partial_z
u_1(0)$ and $\partial_z u_2(h_1)$ as functions of $\overline{u}_i$ and
$h_i$. We clearly see that an expansion of $u_i$ up to order
$\mathcal{O}\big((\alpha+\beta+\delta)^2\big)$ is needed.  We use the
method introduced by Vila \cite{Vila} in the case of a single fluid
layer to write these terms in a closed form. In \cite{Vila}, the wall
stress is chosen proportionnal to the average velocity. One has to
expand both the wall stress and the average velocity up to order $1$
and obtains an expansion of the wall stress with a zeroth order term
proportionnal to the average velocity, the next term depending on the 
fluid height and its time and spatial derivatives. See
\cite{Bresch_Noble} for a mathematical justification of this derivation.\\

\noindent
The situation is more involved for bi-layer flows and several closures
are possible. However, in order to fit with the model in \cite{Vila},
we search an expansion of the fluid stress at the bottom in the form 
$$
\displaystyle
\partial_z u_1(0)=\gamma_1(h_1,h_2)\frac{\overline{u}_1}{h_1}+({\rm h.o.t}), 
$$
\noindent
with $\gamma_1(h_1,0)=3$. The fluid stress at the bottom is given by
\begin{equation}
\displaystyle
\partial_z u_1(0)=\gamma_1(h_1,h_2)\frac{\overline{u}_1}{h_1}+R_1,
\end{equation}
\noindent
with $\gamma_1$ and $R_1$ defined as $\displaystyle\gamma_1=6\frac{h_1+\rho h_2}{2h_1+3\rho h_2}$ and
{\setlength\arraycolsep{1pt}
\begin{eqnarray}
\displaystyle
R_1&=&-\frac{\lambda^2\beta}{2h_1+3\rho h_2}\big(R_{1,1}\partial_x h_1+R_{1,2}\partial_x h_2\big)-\delta c(\rho^2-\rho)\frac{h_1h_2}{2h_1+\rho h_2}\partial_x h_2\nonumber\\
\displaystyle
&&+\frac{\delta h_1h_2}{2h_1+3\rho h_2}\partial_{xxx}\big(\kappa_1\rho h_1+\kappa_2(\rho-1)(h_1+h_2)\big),\nonumber\\
\displaystyle
R_{1,1}&=&\frac{2}{15}h_1^5+\frac{41\rho}{60}h_1^4h_2+\frac{13\rho^2}{10}h_1^3h_2^2+(\frac{3\rho^3}{4}+\frac{\rho^2}{3\nu})h_1^2h_2^3+\frac{\rho^3}{3\nu}h_1h_2^4,\nonumber\\
\displaystyle
R_{1,2}&=&\frac{2\rho}{15}h_1^5+\frac{41\rho^2}{60}h_1^4h_2+(\frac{21\rho^3}{20}+\frac{\rho^2}{4\nu})h_1^3h_2^2+\frac{13\rho^3}{12\nu}h_1^2h_2^3+\frac{\rho^3}{3\nu^2}h_1h_2^4,\nonumber
\end{eqnarray}}
\noindent
Next, we write the fluid stress at the interface as
$$
\partial_z u_2(h_1)\approx 3\frac{\overline{u}_2-u_{int}}{h_2},
$$
\noindent
with $u_{int}=u_1(h_1)=u_2(h_1)$ and expand $u_{int}$ as
$\displaystyle u_{int}=\gamma_2(h_1,h_2)\overline{u}_1+({\rm
  h.o.t})$. The fluid stress at the interface reads
$$
\displaystyle
\partial_z u_2(h_1)=\frac{3}{h_2}(\overline{u}_2-\gamma_2\,\overline{u}_1)+R_2,
$$
\noindent
with $\gamma_2$ and $R_2$ defined as $\displaystyle\gamma_2=3\frac{h_1+2\rho h_2}{2h_1+3\rho h_2}$ and
{\setlength\arraycolsep{1pt}
\begin{eqnarray}
\displaystyle
R_2&=&-\frac{\beta\lambda^2}{h_2(2h_1+3\rho h_2)}\big(R_{2,1}\partial_x h_1+R_{2,2}\partial_x h_2\big)-\delta c(\rho^2-\rho)\frac{3h_1^2\partial_x h_2}{2(2h_1+3\rho h_2)}\nonumber\\
\displaystyle
&&+\frac{3\delta h_1^2}{2(2h_1+3\rho h_2)}\partial_{xxx}\big((\kappa_1\rho+\kappa_2(\rho-1))h_1+\kappa_2(\rho-1)h_2\big)\nonumber\\
\displaystyle
R_{2,1}&=&\frac{1}{20}h_1^6+\frac{2\rho}{5} h_1^5h_2+\frac{11\rho^2}{10}h_1^4h_2^2+(\frac{\rho^2}{2\nu}+\frac{3\rho^3}{4})h_1^3h_2^3\nonumber\\
\displaystyle
&&+(\frac{2\rho^2}{15\nu^2}+\frac{\rho^3}{2\nu})h_1^2h_2^4+\frac{\rho^3}{3\nu^2}h_1h_2^5+\frac{\rho^4}{5\nu^2}h_2^6\nonumber\\
\displaystyle
R_{2,2}&=&\frac{\rho}{20}h_1^6+\frac{2\rho^2}{5}h_1^5h_2+(\frac{\rho^2}{8\nu}+\frac{117\rho^3}{120})h_1^4h_2^2+\frac{5\rho^3}{4\nu}h_1^3h_2^3\nonumber\\
\displaystyle
&&+\frac{19\rho^3}{30\nu^2}h_1^2h_2^4+(\frac{\rho^4}{5\nu^2}+\frac{2\rho^3}{15\nu^3})h_1h_2^5+\frac{\rho^4}{5\nu^3}h_2^6.\nonumber
\end{eqnarray}}

\noindent
Note that we have implicitely used the mass conservation law 
$$\displaystyle \partial_t h_i=-\partial_x q_i^{(0)}+({\rm h.o.t})$$
to transform time derivatives into {\it spatial} derivatives.  
As a result, we obtain a shallow water model for bi-layer flows in a
closed form:
{\setlength\arraycolsep{1pt}
\begin{eqnarray}
\label{mass12}
\displaystyle
\partial_t h_1+\partial_x(h_1\overline{u}_1)&=&0,\quad
\partial_t h_2+\partial_x(h_2\overline{u}_2)=0,\\
\displaystyle
\label{mom1}\partial_t(h_1\overline{u}_1)+\partial_x(h_1\overline{u}_1^2+\frac{ch_1^2}{2F^2})&=&\frac{1}{\varepsilon\,R_e}\big(\lambda\,h_1+3\nu\frac{\overline{u}_2-\gamma_2\overline{u}_1}{h_2}-3\gamma_1\frac{\overline{u}_1}{h_1}\big)+\mathcal{R}_1,\\
\label{mom2}\partial_t(h_2\overline{u}_2)+\partial_x(h_2\overline{u}_2^2+\frac{ch_2^2}{2F^2})&=&\frac{1}{\varepsilon\,R_e}\big(\lambda\,h_2-\frac{3\nu}{\rho}\frac{\overline{u}_2-\gamma_2\overline{u}_1}{h_2}\big)+\mathcal{R}_2,
\end{eqnarray}}
\noindent
$$
\displaystyle
\mathcal{R}_1=-\frac{c}{F^2}h_2\partial_x
h_1+\widetilde{\mathcal{R}}_1,\quad \mathcal{R}_2=-\frac{c}{F^2}h_1\partial_x
h_2+\widetilde{\mathcal{R}}_2,
$$  
where $\widetilde{\mathcal{R}}_i,\,i=1,2$ are only fonctions of $h_i$
and their spatial derivatives. These are corrective terms to the
hydrostatic repartition of pressure within the fluids which are due to
surface tension, buoyancy and inertia. They are written as
{\setlength\arraycolsep{1pt}
\begin{eqnarray}
\displaystyle
\widetilde{\mathcal{R}}_1&=&-\partial_x Q_1+\frac{1}{\varepsilon
  R_e}(\nu
R_2-R_1)+h_1\partial_{xxx}(\kappa_2(h_1+h_2)+\kappa_1\,h_1),\nonumber\\
\displaystyle
\widetilde{\mathcal{R}}_2&=&-\frac{1}{\rho}\partial_x
Q_2-\frac{\nu}{\varepsilon R_e\rho}R_2+\frac{\kappa_2}{\rho}h_2\partial_{xxx}(h_1+h_2).\nonumber
\end{eqnarray}}

\noindent
This system is not in a conservative from and this may lead to some
indetermination in the presence of shocks. One can drop this
indetermination by the use of ``nonconservative paths'' (see the next
section for definitions). 

\noindent
For convenience we rewrite the system in matrix form. Set $W=(h_1,h_2,\bar{u}_1,\bar{u}_2)^T$. We write system (\ref{mass12}, \ref{mom1}, \ref{mom2}) as
\begin{equation}\label{eq:syst1}
\displaystyle
\partial_t W+\partial_x F(W)=B(W)\partial_x W+G(W),
\end{equation}
\noindent
where $F$ is defined as
\[
F\left(W\right)=\left[\begin{array}{c}
q_{1}\\
q_{2}\\
{\displaystyle \frac{q_{1}^{2}}{h_{1}}+\frac{ch_{1}^{2}}{2F^{2}}+\lambda^{2}h_{1}^{3}\left(\frac{h_{1}^{2}}{45}+\frac{\rho h_{2}}{12}\left(h_{1}+\rho h_{2}\right)\right)}\\
{\displaystyle \frac{q_{2}^{2}}{h_{2}}+\frac{ch_{2}^{2}}{2F^{2}}+\frac{\lambda^{2}\rho h_{2}^{5}}{45\nu^{2}}}\end{array}\right]\]
\noindent
whereas $B$ is given by

\[
B\left(W\right)\frac{\partial W}{\partial x}=\left[\begin{array}{cccc}
0 & 0 & 0 & 0\\
0 & 0 & 0 & 0\\
B_{11} & B_{12} & 0 & 0\\
B_{21} & B_{22} & 0 & 0\end{array}\right]\left[\begin{array}{c}
\partial_{x}h_{1}\\
\partial_{x}h_{2}\\
\partial_{x}q_{1}\\
\partial_{x}q_{2}\end{array}\right]\]
with

\[
\begin{array}{cl}
B_{11} & {\displaystyle =-\frac{c}{F^{2}}h_{2}+\frac{\lambda^{2}}{2h_{1}+3\rho h_{2}}R_{1,1}-\frac{\nu\lambda^{2}}{h_{2}\left(2h_{1}+3\rho h_{2}\right)}R_{2,1}}\\
\\B_{12} & {\displaystyle =-\left[\frac{\lambda^{2}}{h_{2}}R_{2,2}+c\left(\rho^{2}-\rho\right)\frac{3h_{1}^{2}}{2F^{2}}\right]\frac{\nu}{\left(2h_{1}+3\rho h_{2}\right)}}\\
 & {\displaystyle +\frac{\lambda^{2}}{2h_{1}+3\rho h_{2}}R_{1,2}+{\displaystyle \frac{1}{F^{2}}}c\left(\rho^{2}-\rho\right)\frac{h_{1}h_{2}}{2h_{1}+\rho h_{2}}}\\
\\B_{21} & {\displaystyle =\frac{\nu\lambda^{2}}{h_{2}\rho\left(2h_{1}+3\rho h_{2}\right)}R_{2,1}}\\
\\B_{22} & {\displaystyle =\left[\frac{\lambda^{2}}{h_{2}}R_{2,2}+c\left(\rho^{2}-\rho\right)\frac{3h_{1}^{2}}{2F^{2}}\right]\frac{\nu}{\rho\left(2h_{1}+3\rho h_{2}\right)}-\frac{c}{F^{2}}h_{1}}.\end{array}\]
The source term $G\left(W\right)$ reads

\[
G\left(W\right)={\displaystyle \frac{1}{\varepsilon R_{e}}}\left[\begin{array}{c}
0\\
0\\
{\displaystyle \lambda h_{1}+3\nu\frac{\bar{u}_{2}-\gamma_{2}\bar{u}_{1}}{h_{2}}-3\gamma_{1}\frac{\bar{u}_{1}}{h_{1}}}\\
{\displaystyle \lambda h_{2}+\frac{3\nu}{\rho}\frac{\bar{u}_{2}-\gamma_{2}\bar{u}_{1}}{h_{2}}}\end{array}\right]\]

\section{\label{sec2} Roll-waves in shallow water equations}

In this section, we prove the existence of roll-waves in bilayer flows
when steady states are unstable. They are defined as piecewise smooth
and spatially periodic travelling waves, entropic solutions to shallow
water equations. For hyperbolic conservation laws, the shocks must
satisfy the Rankine-Hugoniot and Lax shock conditions. In
\cite{Dressler}, these solutions are proved to exist in a single layer 
of fluid modeled by a shallow water system. For general hyperbolic
conservations laws with source terms, there are small amplitude
roll-waves \cite{Noble_rwgs}. We generalize this result to
nonconservative hyperbolic systems in order to deal
with our shallow water model for bi-layer flows. We also carry out direct
numerical simulations to show the existence of large amplitude
roll-waves.

\subsection{Existence of small amplitude roll-waves}

In this section, we consider the problem
\begin{equation}\label{edp_hyp_nc}
\displaystyle
\frac{\partial u}{\partial t}+\mathcal{A}(u)\frac{\partial u}{\partial x}=g(u),\quad x\in\mathbb{R},\:\:t>0.
\end{equation}
\noindent
We assume that system (\ref{edp_hyp_nc}) is strictly hyperbolic in the neighbourhood $\mathcal{V}(\overline{u}_0)$ of a constant solution $u=\overline{u}_0$, meaning that $\mathcal{A}(u)$ has $n$ real distinct eigenvalues $\big(\lambda_k(u)\big)_{k=1...n}$, 
\begin{equation*}
\displaystyle
\lambda_1(u)<...<\lambda_k(u)<...<\lambda_n(u),\quad \forall\, u\in\mathcal{V}(\overline{u}_0).
\end{equation*}
\noindent
We suppose that both $\mathcal{A}$ and $g$ have a power serie
expansion at $u=\overline{u}_0$ with a disk of convergence containing
$B(\overline{u}_0,r)$ for a suitable $r>0$. As in \cite{Dressler}, we search a spatially periodic travelling wave $u(x,t)=U(x-ct)$ with $U$ a $2L$-periodic function with discontinuities at $x_j=(2j+1)L,\;j\in\mathbb{Z}$ which satisfies the differential system
\begin{equation}\label{edo_rw}
\displaystyle
\big(\mathcal{A}(U(x))-c\big)U'=g(U(x)),\quad \forall x\in(-L,\,L).
\end{equation}
\noindent
Next, we formulate conditions for shocks at
$x_j=(2j+1)L,\;j\in\mathbb{Z}$. For that purpose, we need to define the so-called ``family of paths'' \cite{MaFloMu}. These paths were introduced to give a rigorous definition of nonconservative products in hyperbolic systems. A family of paths $\Phi$ in $\Omega\subset\mathbb{R}^n$ is a locally Lipschitz map
$\displaystyle \Phi: [0,\;1]\times\Omega\times\Omega\to \Omega,$ such that
\begin{itemize}
\item $\Phi(0,u_l,u_r)=u_l$ and $\phi(1,u_l,u_r)=u_r$, for any $u_l,u_r\in\Omega$;
\item for any bounded subset $\mathcal{O}\subset\Omega$, there exists a constant $k$ such that $\displaystyle\left|\frac{\partial\Phi}{\partial s}(s,u_l,u_r)\right|\leq k|u_l-u_r|$, for all $u_r,u_l\in\mathcal{O}$ and almost every $s\in[0,\:1]$.
\item for any bounded subset $\mathcal{O}\subset\Omega$, there exists a constant $K$ such that
\begin{equation*}
\displaystyle
\left|\frac{\partial\Phi}{\partial s}(s,u_l^1,u_r^1)-\frac{\partial\Phi}{\partial s}(s,u_l^2,u_r^2)\right|\leq K\big(|u_l^1-u_l^2|+|u_r^1-u_r^2|\big),
\end{equation*} 
for all $u_r^1,u_r^2,u_l^1,u_l^2\in\mathcal{O}$ and almost every $s\in[0,\,1]$. 
\end{itemize} 
\noindent
When such a family has been chosen, one can define generalized Rankine
Hugoniot jump condition across a discontinuity with speed $\xi$
\begin{equation*}
\displaystyle
\int_0^{1}\Big(\xi Id-\mathcal{A}\big(\Phi(s,u^-,u^+)\big)\Big)\frac{\partial\Phi}{\partial s}(s,u^-,u^+)\,ds=0
\end{equation*}
\noindent
where $u^-, u^+$ are the left and right limits at the
discontinuity. For the problem of roll-waves, this can be written as a
nonlinear boundary condition, setting $u^-=U(L)$ and $u^+=U(-L)$:
\begin{equation}\label{rh_rw}
\displaystyle
\int_0^{1}\Big(c Id-\mathcal{A}\big(\Phi(s,U(L),u(-L))\big)\Big)\frac{\partial\Phi}{\partial s}(s,U(L),U(-L))\,ds=0.
\end{equation}
\noindent
This Rankine-Hugoniot condition is completed with a Lax shock condition
\begin{equation}\label{lax_rw}
\displaystyle
\lambda_k(U(-L))<c<\lambda_k(U(L)),\quad \lambda_{k-1}(U(L))<c<\lambda_{k+1}(U(-L)),
\end{equation}
 \noindent
 for some $k$, $1\leq k\leq n$. Herein, a roll-wave is a solution of the so called ``roll-wave problem'' (\ref{edo_rw},\ref{rh_rw},\ref{lax_rw}).\\

\noindent 
Let us fix $k$, $1\leq k\leq n$: we denote $r_k(u)$ the eigenvector of
$\mathcal{A}(u)$ associated to the eigenvalue $\lambda_k(u)$ and we
assume that the characteristic field $r_k(u)$ is genuinely nonlinear $\displaystyle
\nabla \lambda_k(u).r_k(u)\neq 0,\quad\forall\,u\in\mathcal{V}(\overline{u}_0).$
\noindent
We define $\Pi_k(u)$ as the projection onto ${\rm
  Ker}(\mathcal{A}(u)-\lambda_k(u)Id)$ with respect to ${\rm
  Im}(\mathcal{A}(u)-\lambda_k(u)Id)$. The eigenvalue is isolated so
that $u\mapsto \Pi_k(u)$ is $C^1$. ${\rm
  Ker}(\mathcal{A}(u)-\lambda_k(u)Id)$ is one dimensional and we
identify, for any $v\in\mathbb{R}^n$, $\Pi_k(u)v$ to a real
number. Finally, we will suppose that $\Phi$ is analytic in each
variables in order to simplify the discussion: this hypothesis is
clearly statisfied for straight lines, a natural choice in numerical
schemes. Let us prove the existence of small amplitude roll-waves.
\begin{theorem}
Assume that
\begin{equation}
\Pi_k(\overline{u}_0)D\mathcal{A}(\overline{u}_0).r_k(\overline{u}_0).r_k(\overline{u}_0)\neq 0,\quad \Pi_k(\overline{u}_0)dg(\overline{u}_0).r_k(\overline{u}_0)\neq 0,
\end{equation}
\begin{equation}\label{exrw0}
\displaystyle
\frac{\Pi_k(\overline{u}_0)dg(\overline{u}_0).r_k(\overline{u}_0)}{\Pi_k(\overline{u}_0)D\mathcal{A}(\overline{u}_0).r_k(\overline{u}_0).r_k(\overline{u}_0)}\nabla\lambda_k(\overline{u}_0).r_k(\overline{u}_0)>0,
\end{equation}
and $\displaystyle dg(\overline{u}_0):\mathbb{R}^n\to\mathbb{R}^{n}$ is invertible. Then there exists a family of {\it small amplitude} roll-waves solutions of (\ref{edo_rw},\ref{rh_rw},\ref{lax_rw}) parametrized by wavelength.
\end{theorem}

\noindent
{\bf Proof.} We will prove that the solutions to
(\ref{edo_rw},\ref{rh_rw},\ref{lax_rw}) are zeros of a submersion
between suitable functional spaces. First, let us recall the construction of a formal roll-wave. Set
$L=2\eta \tau$. We search a roll-wave in the form $u(x,t)=\overline{u}+\eta v(\frac{x-ct}{\eta\tau})$, $\eta\ll 1$. The system (\ref{edo_rw},\ref{rh_rw},\ref{lax_rw}) reads
\begin{equation}\label{edo_rw_1}
\displaystyle
\big(\mathcal{A}(\overline{u}+\eta\,v(x))-c\big)v'(x)=\tau\,g(\overline{u}+\eta v(x)),\quad\forall\,x\in(-1,\,1),
\end{equation} 
\begin{equation}\label{rh_rw_1}
\displaystyle
\displaystyle
\int_0^{1}\Big(c Id-\mathcal{A}\big(\Phi(s,\overline{u}+\eta v(1),\overline{u}+\eta v(-1))\big)\Big)\frac{\partial\Phi}{\partial s}(s,\overline{u}+\eta v(1),\overline{u}+\eta v(-1))\,ds=0.
\end{equation}
As $\eta\to 0$, the Lax shock conditions are 
\begin{equation}\label{lax_rw_1}
\displaystyle
\lambda_k(\overline{u}+\eta v(-1))<c<\lambda_k(\overline{u}+\eta v(1)).
\end{equation}
\noindent
Letting $\eta\to 0$ in (\ref{rh_rw_1}) yields $\displaystyle
\mathcal{A}(\overline{u})\big(v(1)-v(-1)\big)=c\big(v(1)-v(-1)\big).$
Necessarily $c=\lambda_k(\overline{u})$ and $v(1)-v(-1)$ is an
eigenvector of $\mathcal{A}(\overline{u})$ associated to
$\lambda_k(\overline{u})$. Let us search
$v(x)=\alpha(x)r_k(\overline{u})$: the Rankine Hugoniot condition is
then satisfied. If $\overline{u}=\overline{u}_0$ a zero of $g$, then
dividing (\ref{edo_rw_1}) by $\eta$ and letting $\eta\to 0$ yields
\begin{equation}
\displaystyle
D\mathcal{A}(\overline{u}_0).r_k(\overline{u}_0).r_k(\overline{u}_0)\,\alpha(x)\alpha'(x)=dg(\overline{u}_0).r_k(\overline{u}_0)\alpha(x).
\end{equation} 
\noindent
Then, a projection onto ${\rm
  Ker}(\mathcal{A}(\overline{u}_0)-\lambda_k(\overline{u}_0))$ yields
$\alpha(x)=\Gamma\, x$. As $\eta\to 0$, the Lax shock conditions are
\begin{equation}
\displaystyle
\nabla\lambda_k(\overline{u}_0).r_k(\overline{u}_0)\alpha(-1)<0<\nabla\lambda_k(\overline{u}_0).r_k(\overline{u}_0)\alpha(1).
\end{equation}
\noindent
This relation is clearly satisfied under the assumption (\ref{exrw0})
and the construction of a formal roll-wave is complete. In order to
prove the existence of roll-waves close to this formal solution, we introduce
{\setlength\arraycolsep{1pt}
\begin{eqnarray*}
\mathcal{F}_{\varepsilon,\tau}&:&\mathbb{X}\times\mathcal{V}(\overline{u}_0)\to \mathbb{Y}\times\mathbb{R}^n\nonumber\\
\displaystyle
\mathcal{F}_{\eta,\tau}(v,\overline{u})_1&=&\Pi_k(\overline{u})\Big(\frac{\mathcal{A}(\overline{u}+\eta v)-\mathcal{A}(\overline{u})}{\eta}-\tau\frac{g(\overline{u}+\eta)-<g(\overline{u}+\eta v)>}{\eta}\Big)\nonumber\\
\displaystyle
&+&(Id-\Pi_k(\overline{u}))\Big(\big(\mathcal{A}(\overline{u}+\eta v)-\lambda_k(\overline{u})\big)v'\Big)\nonumber\\
\displaystyle
&+&\tau(Id-\Pi_k(\overline{u}))\big(<g(\overline{u}+\eta v)>-g(\overline{u}+\eta v)\big),\nonumber\\
\displaystyle
\mathcal{F}_{\eta,\tau}(v,\overline{u})_2&=&\int_{-1}^{1}g(\overline{u}+\eta\,v(x))dx+\eta\int_{-1}^2\frac{\mathcal{A}(\overline{u})-\mathcal{A}(\overline{u}+\eta v(x))}{\eta}v'(x)dx\nonumber\\
&+&\eta^{-1}\int_0^1\mathcal{A}\big(\Phi(s,\overline{u}+\eta v(\pm 1))-\mathcal{A}(\overline{u})\big)\frac{\partial\Phi}{\partial s}(s,\overline{u}+\eta v(\pm 1))ds,
\end{eqnarray*}}
\noindent
with $<g(\overline{u}+\eta)>=\frac{1}{2}\mathcal{F}_{\eta,\tau}(\overline{u},v)_2$. The functional spaces $\mathbb{X},\mathbb{Y}$ are defined as
\begin{equation*}
\begin{array}{lll}
\displaystyle
\mathbb{X}_0=\left\{f\in C^1(-1,1)/f(x)=\sum_{n\geq 0}a_n\,x^n,\:\sum_{n\geq 0}(n+1)|a_n|<\infty\right\},\\
\displaystyle
\mathbb{X}=\left\{f\in\mathbb{X}_0\,/\,(1-\Pi_k(\overline{u}_0))\int_{-1}^1f(x)dx=0\right\},\nonumber\\
\displaystyle
\mathbb{Y}=\left\{f\in C^1(-1,1)/\;f(x)=\sum_{n\geq 0}a_nx^n,\:\sum_{n\geq 0}|a_n|<\infty\right\}.
\end{array}
\end{equation*}
\noindent
The operator $\mathcal{F}_{\eta,\tau}$ is well defined and $C^1$. It
is clear that a zero $(\overline{u},v)$ of $\mathcal{F}_{\eta,\tau}$
corresponds to a roll-wave (see \cite{Noble_rwgs} for more
details). Let us fix $\tau_0>0$. As $\eta\to 0$, it is easily seen
that a zero $(\overline{u},v)$ of $\mathcal{F}_{0,\tau_0}$ satisfies
\begin{equation}
\begin{array}{lll}
\displaystyle
\Pi_k(\overline{u})D\mathcal{A}(\overline{u}).v(x).v'(x)-\Pi_k(\overline{u})dg(\overline{u})\big(v(x)-\frac{1}{2}\int_{-1}^1 v(s)ds\big),\\
\displaystyle
(1-\Pi_k(\overline{u}))\Big(\mathcal{A}(\overline{u})-\lambda_k(\overline{u})\Big)v'(x)=0,\quad\forall\,x\in(-1,\,1),\\
\displaystyle
g(\overline{u})=0.
\end{array}
\end{equation}
The roll-wave
$\big(\overline{u}_0,v(x)=\alpha(x)r_k(\overline{u}_0)\big)$ is
clearly a zero of $\mathcal{F}_{0,\tau_0}$. The end of the proof is
similar to the conservative case. One prove that
$D\mathcal{F}_{0,\tau_0}(\overline{u}_0,\alpha(x)r_k(\overline{u}_0))$
is invertible and the implicit function theorem applies: for
$0<\eta\ll 1$ and $\tau\approx\tau_0$, there exist a unique zero of
$\mathcal{F}_{\eta,\tau}$ which is close to the formal roll-wave. If
condition (\ref{exrw0}) is satisfied, the roll-wave satisfies Lax
shock condition (\ref{lax_rw_1}) for $\eta>0$ sufficiently small and
the proof is complete. $\Box$\\

\noindent
A slight modification of this argument enables us to deal
with``physical'' source terms (which is the case for bi-layer
flows). Suppose that $g$ has the particular form
$g(u)=\big(0_{\mathbb{R}^p},h(u)\big)$ and
$dh(\overline{u}_0):\mathbb{R}^n\to\mathbb{R}^{n-p}$ is onto. The
construction of the formal roll-wave is the same. Then one can prove
that for $0<\eta\ll1$ and $\tau\approx\tau_0$, there exists a family
of roll-waves solutions that belongs to a $p+1$-dimensional manifold. In this case, $\mathcal{F}_{\eta,\tau}$ is a submersion at the point corresponding to the formal roll-wave.\\

\subsection{Numerical simulations}

In this section, we investigate numerically the existence of
roll-waves through direct numerical simulations of the shallow water
equations (\ref{mass12}, \ref{mom1}, \ref{mom2}). We consider the
case where all the interfaces are unstable (these
latter solutions are the bilayer counterpart of regular roll-waves
into a single fluid layer). 

\subsubsection{Numerical scheme}

\noindent
We use a classical upwind difference scheme as described in \cite{Leer}. We assume
$x\in\left[0,L\right]$ and integrate (\ref{eq:syst1}) on the time
interval $\left[0,T\right]$. System \eqref{eq:syst1} is strictly
hyperbolic provided that the eigenvalues of $M\left(W\right)=A\left(W\right)-B\left(W\right)$ are real and distincts. Here $A\left(W\right)$ denotes the Jacobian matrix of $F$:
\[
A={\displaystyle \frac{\partial F}{\partial W}}=\left[\begin{array}{cccc}
0 & 0 & 1 & 0\\
0 & 0 & 0 & 1\\
A_{31} & A_{32} & {\displaystyle \frac{2q_{1}}{h_{1}}} & 0\\
0 & A_{42} & 0 & {\displaystyle \frac{2q_{2}}{h_{2}}}\end{array}\right]\]
with (recall that the nondimensional numbers $\beta, \delta, \lambda$ are defined as $\beta=\varepsilon R_{e}$ , $\delta={\displaystyle \frac{\varepsilon R_{e}}{F^{2}}}$ and
$\lambda={\displaystyle \frac{R_{e}}{F^{2}}}\sin\theta$) :

\[
\begin{array}{l}
{\displaystyle A_{31}=-\frac{q_{1}^{2}}{h_{1}^{2}}+\frac{ch_{1}}{F^{2}}+3\lambda^{2}h_{1}^{2}\left(\frac{h_{1}^{2}}{45}+\frac{\rho h_{2}}{12}\left(h_{1}+\rho h_{2}\right)\right)+\lambda^{2}h_{1}^{3}\left(\frac{2h_{1}}{45}+\frac{\rho h_{2}}{12}\right)}\\
{\displaystyle A_{32}=\lambda^{2}h_{1}^{3}\left[\frac{\rho}{12}\left(h_{1}+\rho h_{2}\right)+\frac{\rho^{2}h_{2}}{12}\right]=\rho\lambda^{2}h_{1}^{3}\left[\frac{h_{1}}{12}+\frac{\rho h_{2}}{6}\right]}\\
{\displaystyle A_{42}=-{\displaystyle \frac{q_{2}^{2}}{h_{2}^{2}}+\frac{ch_{2}}{F^{2}}+\frac{5\lambda^{2}\rho h_{2}^{4}}{45\nu^{2}}}}.\end{array}\]
\noindent
As a consequence the matrix $M$ is given by

\begin{equation}
M\left(W\right)=A\left(W\right)-B\left(W\right)=\left[\begin{array}{cccc}
0 & 0 & 1 & 0\\
0 & 0 & 0 & 1\\
M_{31} & M_{32} & {\displaystyle \frac{2q_{1}}{h_{1}}} & 0\\
M_{41} & M_{42} & 0 & {\displaystyle \frac{2q_{2}}{h_{2}}}\end{array}\right]\label{eq:syst2}\end{equation}
and 

\[
\begin{array}{rl}
M_{31}= & {\displaystyle -\frac{q_{1}^{2}}{h_{1}^{2}}+\frac{ch_{1}}{F^{2}}+3\lambda^{2}h_{1}^{2}\left(\frac{h_{1}^{2}}{45}+\frac{\rho h_{2}}{12}\left(h_{1}+\rho h_{2}\right)\right)+\lambda^{2}h_{1}^{3}\left(\frac{2h_{1}}{45}+\frac{\rho h_{2}}{12}\right)}\\
 & {\displaystyle +\frac{c}{F^{2}}h_{2}-\frac{\lambda^{2}}{2h_{1}+3\rho h_{2}}R_{1,1}+\frac{\nu\lambda^{2}}{h_{2}\left(2h_{1}+3\rho h_{2}\right)}R_{2,1}}\\
\\M_{32}= & {\displaystyle \rho\lambda^{2}h_{1}^{3}\left[\frac{h_{1}}{12}+\frac{\rho h_{2}}{6}\right]+\left[\frac{\lambda^{2}}{h_{2}}R_{2,2}+c\left(\rho^{2}-\rho\right)\frac{3h_{1}^{2}}{2F^{2}}\right]\frac{\nu}{\left(2h_{1}+3\rho h_{2}\right)}}\\
 & {\displaystyle -\frac{\lambda^{2}}{2h_{1}+3\rho h_{2}}R_{1,2}-{\displaystyle \frac{1}{F^{2}}}c\left(\rho^{2}-\rho\right)\frac{h_{1}h_{2}}{2h_{1}+\rho h_{2}}}\end{array}\]

\[
\begin{array}{rl}
M_{41}= & {\displaystyle -\frac{\nu\lambda^{2}}{h_{2}\rho\left(2h_{1}+3\rho h_{2}\right)}R_{2,1}}\\
\\M_{42}= & -{\displaystyle \frac{q_{2}^{2}}{h_{2}^{2}}+\frac{ch_{2}}{F^{2}}+\frac{5\lambda^{2}\rho h_{2}^{4}}{45\nu^{2}}+\frac{c}{F^{2}}h_{1}}\\
 & {\displaystyle -\left[\frac{\lambda^{2}}{h_{2}}R_{2,2}+c\left(\rho^{2}-\rho\right)\frac{3h_{1}^{2}}{2F^{2}}\right]\frac{\nu}{\rho\left(2h_{1}+3\rho h_{2}\right)}}\end{array}\]
We approximate system  \eqref{eq:syst1} by the regularized system:

{\setlength\arraycolsep{1pt}
\begin{eqnarray}
\displaystyle
\frac{\partial W}{\partial t}+\frac{\partial}{\partial x}F\left(W\right)&=&G\left(W\right)+B\left(W\right)\frac{\partial W}{\partial x}\nonumber\\
&+&\frac{\partial}{\partial x}\left(\frac{\Delta x}{2}\mathcal{D}\left(W\right)\frac{\partial W}{\partial x}\right).
\end{eqnarray}}

\noindent
where ${\displaystyle \frac{\Delta x}{2}}\partial_{x}\left(\mathcal{D}\left(W\right)\partial_{x}W\right)$ represents a numerical diffusion introduced by the scheme. The diffusion matrix $\mathcal D \left(W\right)$ must take into account that the effective transport term in \eqref{eq:syst1} is $M\left(W\right)\partial_xW$, where $M$ is given by \eqref{eq:syst2}. We propose to discretize the nonconservative product using the relation $BW_x=\left(BW\right)_x-B_xW$. We use a two-stage second order time scheme.


\noindent
The numerical scheme is then written as
{\setlength\arraycolsep{1pt}
\begin{eqnarray}
\displaystyle
W^{n+1}_i&=& W_i^n+\frac{\Delta t}{2} \left( f\left(W_i^n\right) + f\left(W_i^{n+\frac{1}{2}} \right) \right)\nonumber \\
W^{n+\frac{1}{2}}_i&=& W_i^n+\Delta t f\left(W_i^n\right)\nonumber \\
f\left(W_i^n\right)&=&G\left(W_i^n\right)-\frac{\phi_i^n-\phi_{i-1}^n}{\Delta x} \nonumber \\
&& +\frac{B\left(\widetilde W_{i+\frac{1}{2}}^n \right) \widetilde W^n_{i+\frac{1}{2}} - B\left(\widetilde W^n_{i-\frac{1}{2}} \right) \widetilde W^n_{i-\frac{1}{2}}  }{2\Delta x} \nonumber \\
&& -\frac{ B\left(\widetilde W^n_{i+\frac{1}{2}} \right) - B\left(\widetilde W^n_{i-\frac{1}{2}} \right) }  {\Delta x}\;W^n_i \nonumber ,
\end{eqnarray}}

\noindent
where $\widetilde W_i^n$ is the second order MUSCL reconstructed state of $W_i^n$ using classical flux limiter function \textsl{minmod} as described in \cite{VilaMUSCL}, $\widetilde W^n_{i+\frac{1}{2}}=\frac{W_{i+1}^n+W_i^n}{2}$ is an intermediate state between $W_i^n$ and $W_{i+1}^n$, and the numerical flux $\phi^n_i$ is given by



\[
\displaystyle
\phi_i^n=F_{C}\left(\widetilde W_i^n,\widetilde W^n_{i+1}\right)-\frac{\Delta x}{2} D\left(\widetilde W^n_i,\widetilde W^n_{i+1}\right)\frac{\widetilde W^n_{i+1}-\widetilde W^n_{i}}{\Delta x}
\]
\noindent
where $F_{C}$ and $D$ are respectively approximations of $F$ and $\mathcal{D}$ at $x=x_{i+\frac{1}{2}}$ :
{\setlength\arraycolsep{1pt}
\begin{eqnarray}
\displaystyle
F_{C}\left(U,V\right)&=&\frac{F\left(U\right)+F\left(V\right)}{2}\nonumber \\
D\left(U,V\right)&=&X\left|\Lambda\right| X^{-1} \nonumber
\end{eqnarray}}

\noindent
with $\Lambda$ the matrix of the eigenvalues of $M\left(\frac{U+V}{2}\right)$ and $X$ the matrix defined by its eigenvectors. Note that we note $\left(d_i\right)_{i=1,\dots,4}$ the eigenvalues of $D$, the CFL condition is then given by :
$$
\frac{\Delta t}{\Delta x}\max_{i=1,\dots,4} d_i\leq1
$$
\subsubsection{Numerical results}
 
We carry out numerical simulations when the steady states are
unstable. 
We have fixed: $\nu=0,9$ and
$\rho=0,5$ (see figure \ref{fig3}) and set $\theta=\pi/4$ so that $R_{e}=f$ and $F^{2}=\sqrt{2}f/6$. The aspect ratio $\varepsilon$ is $0,01\ll 1$. 
%
We made numerical simulations for $f=0,5$ (all interfaces
are unstable). 
The initial condition is the steady state $\bar{W}=(1,\, 1,\, 1.75,\, 3.56)^T$
perturbed in the direction of the most unstable eigenvector.\\


\noindent
Recall that we fix $f=0.5$ and $\big(R_{e}=f, F^{2}={\displaystyle \frac{\sqrt{2}f}{6}}\big)$. Let us build  the initial condition: the unique unstable eigenvalue of matrix $\partial_WS-2\pi M$ is $\lambda_{2}=13.4444+3.075i$ and the associated eigenvector is 
\[
\begin{array}{l}
\Lambda_{2}=\left[\begin{array}{c}
 0.0436415-0.0321769i\\
-0.0916431-0.400677i\\
 0.047492+0.109129i\\
 0.902194+0i\\
\end{array}\right]=\Phi_{2}^{re}+i\Phi_{2}^{im}\end{array}.\]
We start the numerical simulations with the initial condition
\[
W_{init}=\bar{W}+5.10^{-3}\left(\cos\left(2\pi x\right)\Phi_{2}^{re}-\sin\left(2\pi x\right)\Phi_{2}^{im}\right).\]
\noindent
We took $250$ points for one period in the spatial mesh. At $t=0$,  the fluid interface and the free surface are periodic with the same period but different amplitude. As we used a $5\promille$ perturbation, interfaces are close to steady state. Note that the scale is different for internal and free surface wave: on the left is the scale for  $h_{1}+h_{2}$ whereas the scale on the right corresponds to the free surface $h_{1}$. 
\begin{figure}[h!]
\begin{center}
\fbox{\includegraphics[scale=0.80]{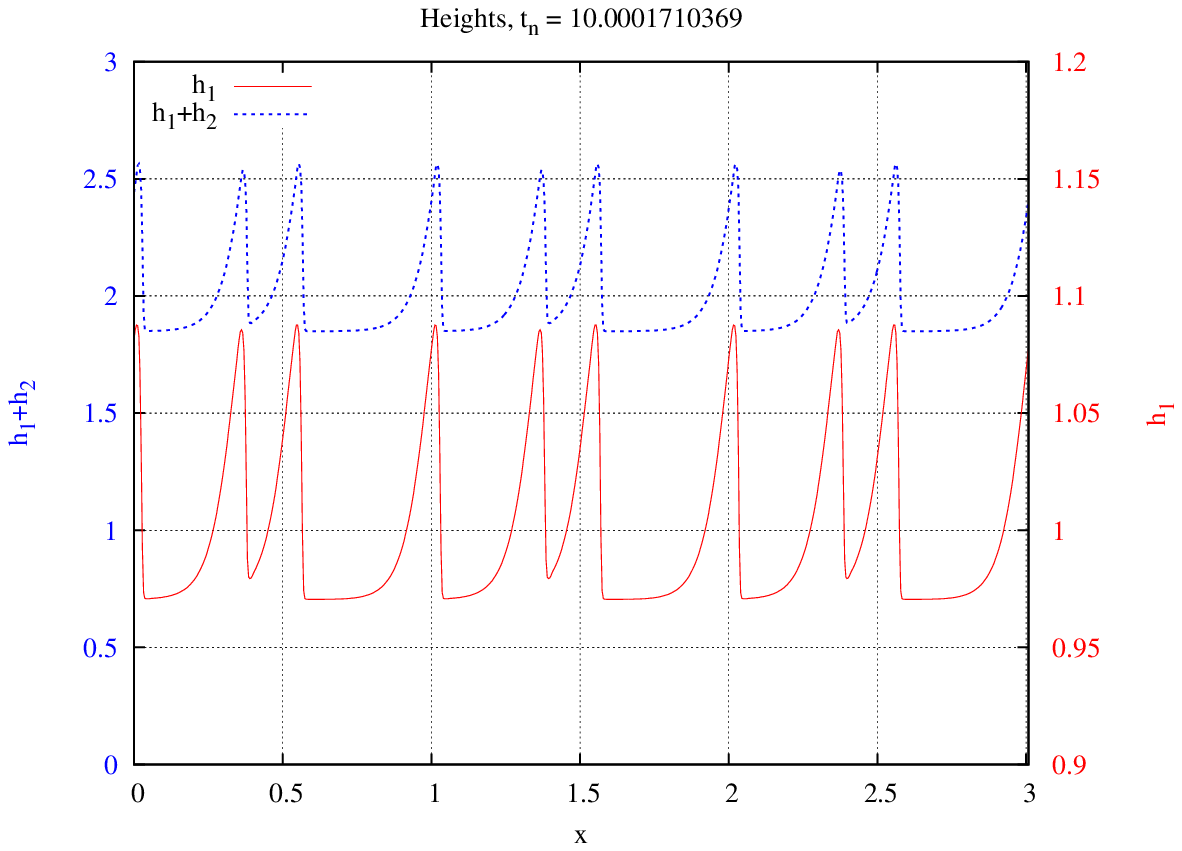}}
\end{center}
\caption{Fluid heights at time $t=10$}
\end{figure}
\noindent
We clearly see the formation of roll-waves both at the free surface
and at the interface and that they are in phase. We have computed the
spatial Fourier transform of this signal: it is composed of $60$ different
modes and the first $20$ modes are the most relevant.
We also plotted in picture \ref{Flo:amplitudes1} the time evolution of the first two Fourier modes : for $t\in[0,\, 1.2]$, the amplitudes of both interface does not vary much and for $t\in[1.2,\, 2]$, there is creation of roll-waves which then stabilize.

\begin{figure}[h]
\begin{center}
\fbox{\includegraphics[scale=0.86]{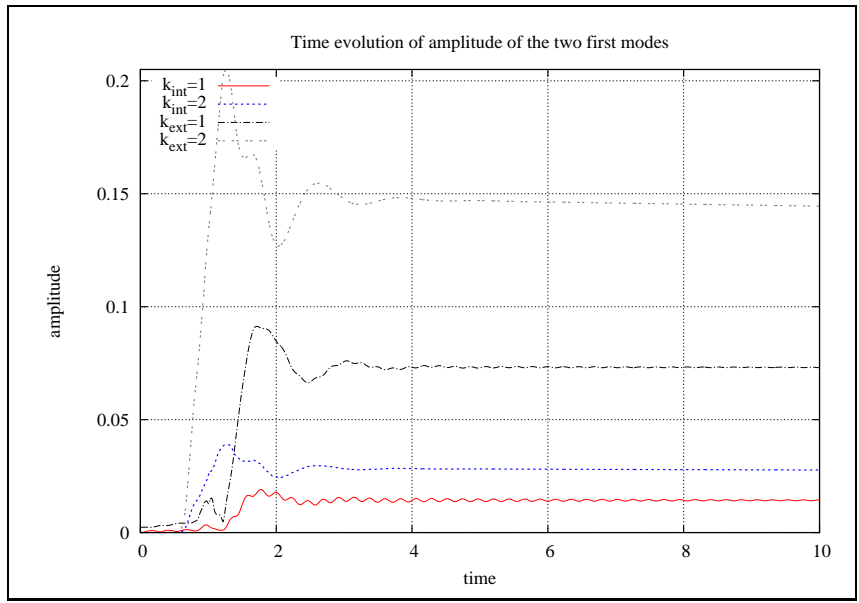}}
\end{center}
\caption{Time evolution of the first two Fourier modes}
\label{Flo:amplitudes1}
\end{figure}

\section{Conclusion}

In this paper, we have obtained consistent shallow water equations
for bi-layer flows from the Navier Stokes equations in the presence of
capillarity. As a byproduct, we carry out a complete spectral
stability analysis of bi-layer flows {\it in the low frequency
  regime}. We proved that this system is a generalization of the
system of Kuramoto Sivashinsky equations derived in \cite{kliak} and
that it is useful to describe nonlinear waves {\it of arbitrary
  amplitude} in bi layer flows. Numerical simulations then confirm the
existence of well known hydrodynamic instabilities, so called
roll-waves, which could be localized on the fluid interface or on both
interfaces.\\

\noindent
This system of shallow water equations is a hyperbolic system in a non
conservative form, a common property in shallow water systems
describing bi-layer flows. Therefore, there is non uniqueness in the
definition of shocks. One possibility would be to derive higher order
shallow water models with a vanishing viscosity: the physical viscous
term would then select the ``physical'' jump conditions. For
applications purposes, it would be also of interest to derive bi-layer
models for {\it non Newtonian} fluids in the spirit of \cite{BNNV},
where consistant shallow water equations for single thin layers of Bingham and 
power law fluids.

\end{document}